\documentclass[twocolumn,showpacs,preprintnumbers,superscriptaddress,pre]{revtex4}
\usepackage{graphicx,bm,times,url}
\usepackage{amsfonts}
\usepackage{amsmath}
\graphicspath{{./fig/}{./png/}}
\newcommand{\BoldVec}[1]{\mathchoice%
  {\mbox{\boldmath $\displaystyle     #1$}}%
  {\mbox{\boldmath $\textstyle        #1$}}%
  {\mbox{\boldmath $\scriptstyle      #1$}}%
  {\mbox{\boldmath $\scriptscriptstyle#1$}}%
}
\newcommand{\EQ}{\begin{equation}}
\newcommand{\EN}{\end{equation}}
\newcommand{\EQA}{\begin{eqnarray}}
\newcommand{\ENA}{\end{eqnarray}}

\newcommand{\EEq}[1]{Equation~(\ref{#1})}
\newcommand{\Eq}[1]{Eq.~(\ref{#1})}
\newcommand{\Eqs}[2]{Eqs.~(\ref{#1}) and~(\ref{#2})}

\newcommand{\Sec}[1]{Sec.~\ref{#1}}

\newcommand{\Fig}[1]{Fig.~\ref{#1}}

\newcommand{\bra}[1]{\langle #1\rangle}

\newcommand{\mean}[1]{\overline #1}
\newcommand{\meanB}{\overline{B}}

\newcommand{\meanBB}{\overline{\bm{B}}}

\newcommand{\meanJJ}{\overline{\mbox{\boldmath $J$}}}

\newcommand{\kf}{\mbox{$k_{\rm f}$}}
\newcommand{\kom}{\mbox{$k_\omega$}}

\newcommand{\xx}{\BoldVec{x}{}}

\newcommand{\uu}{\BoldVec{u} {}}

\newcommand{\UU}{\BoldVec{U} {}}

\newcommand{\bb}{\BoldVec{b} {}}

\newcommand{\BB}{\BoldVec{B} {}}

\newcommand{\AAA}{\BoldVec{A} {}}

\newcommand{\eee}{\BoldVec{e} {}}
\newcommand{\jj}{\BoldVec{j} {}}
\newcommand{\JJ}{\BoldVec{J} {}}

\newcommand{\ff}{\BoldVec{f} {}}

\newcommand{\FF}{\BoldVec{F} {}}

\newcommand{\kk}{\BoldVec{k} {}}

\newcommand{\nab}{\BoldVec{\nabla} {}}

\newcommand{\oo}{\BoldVec{\omega} {}}

\newcommand{\RRRR}{\bm{\mathsf{R}}}
\newcommand{\SSSS}{\bm{\mathsf{S}}}

\newcommand{\ii}{{\rm i}}

\newcommand{\DD}{{\rm D} {}}
\newcommand{\dd}{{\rm d} {}}

\def\Pm{\mbox{\rm Pr}_{\rm M}}
\def\Rm{\mbox{\rm Re}_{\rm M}}

\def\Rey{\mbox{\rm Re}}

\def\cs{c_{\rm s}}
\def\kf{k_{\rm f}}

\def\urms{u_{\rm rms}}

\def\orms{\omega_{\rm rms}}
\def\epsfcrit{\epsilon_{\rm f}^{\rm crit}}
\def\epsf{\epsilon_{\rm f}}
\def\epsm{\epsilon_{\rm m}}
\def\etatz{\eta_{\rm t0}}
\def\etat{\eta_{\rm t}}
\def\etaT{\eta_{\rm T}}
\def\Beq{B_{\rm eq}}
\def\Cacrit{C_{\alpha}^{\rm crit}}

\def\Ca{C_{\alpha}}

\newcommand{\Mm}{\,{\rm Mm}}

\newcommand{\ie}{i.e.\ }

\usepackage{color}
\def\blue{\textcolor{black}}

\def\epsf{\epsilon_{\rm f}}

\begin{document}
\preprint{NORDITA 2012-62}

\title{The kinetic helicity needed to drive large-scale dynamos}

\author{Simon Candelaresi}
\affiliation{NORDITA, KTH Royal Institute of Technology and Stockholm University,
Roslagstullsbacken 23, SE-10691 Stockholm, Sweden}
\affiliation{Department of Astronomy, AlbaNova University Center,
Stockholm University, SE-10691 Stockholm, Sweden}

\author{Axel Brandenburg}
\affiliation{NORDITA, KTH Royal Institute of Technology and Stockholm University,
Roslagstullsbacken 23, SE-10691 Stockholm, Sweden}
\affiliation{Department of Astronomy, AlbaNova University Center,
Stockholm University, SE-10691 Stockholm, Sweden}

\date{\today,~ $ $Revision: 1.99 $ $}

\begin{abstract}
Magnetic field generation on scales large compared with the scale of
the turbulent eddies is known to be possible via the so-called $\alpha$
effect when the turbulence is helical and if the domain is large enough
for the $\alpha$ effect to dominate over turbulent diffusion.
Using three-dimensional turbulence simulations, we show that the energy
of the resulting mean magnetic field of the saturated state
increases linearly with the product
of normalized helicity and the ratio of domain scale to eddy scale,
provided this product exceeds a critical value of around unity.
This implies that large-scale dynamo action commences when the normalized
helicity is larger than the inverse scale ratio.
\blue{
Our results show that the emergence of small-scale dynamo action does
not have any noticeable effect on the large-scale dynamo.
}
Recent findings by Pietarila Graham et al.\ (2012, Phys.\ Rev.\ E85, 066406)
of a smaller minimal helicity may be an artifact due to the onset
of small-scale dynamo action at large magnetic Reynolds numbers.
However, the onset of large-scale dynamo action is difficult to establish when
the kinetic helicity is small.
Instead of random forcing, they used an ABC-flow with time-dependent phases.
We show that such dynamos saturate prematurely in a way that is reminiscent
of inhomogeneous dynamos with internal magnetic helicity fluxes.
Furthermore, even for very low fractional helicities, such dynamos
display large-scale fields that change direction, which is uncharacteristic
of turbulent dynamos.
\end{abstract}

\pacs{47.65.Md, 07.55.Db, 95.30.Qd, 96.60.Hv}
\keywords{large-scale dynamos; alpha effect; helicity}

\maketitle

\section{Introduction}

The origin of magnetic fields in astrophysical bodies like the Earth, the Sun
and galaxies is studied in the field of dynamo theory.
The temporal variation and strength of those fields rules out a primordial
origin, through which the magnetic field would have been created
in the early Universe.
For magnetic fields with energies of the equipartition value, \ie the
turbulent kinetic energy of the medium, the primordial hypothesis explains
their strength after creation, but falls short of explaining how the field
outlives billions of years of resistive decay \cite{Parker1979book}.

In dynamo theory, astrophysical plasmas are considered sufficiently well
conducting fluids where the inertia of the charge-carrying particles
can be neglected.
In this approximation the equations of magnetohydrodynamics (MHD) provide
an adequate model of the medium.
In this framework it has been studied under which conditions magnetic fields
of equipartition strength and scales larger than the turbulent motions are
created and sustained \cite{MoffattBook1978}.

A successful theoretical model describing the dynamo's behavior
is the mean-field theory.
It relates the small-scale turbulent motions to the mean magnetic field
via the so-called $\alpha$ effect, which provides the energy input via
helical turbulent forcing.
During the kinematic phase, \ie negligible back reaction of the magnetic field
on the fluid, the $\alpha$ effect gives a positive feedback on the large-scale
magnetic field, which results in its exponential growth.
Already the consideration of the kinematic MHD equations with negligible Lorentz force
sheds light on the growth rate of the different modes of the magnetic field
during the kinematic phase.
In the kinematic phase the growth rate $\lambda$ at wave number $k$ is given by
\cite{MoffattBook1978}
\EQ \label{eq: lambda kinetic}
\lambda  = \alpha k - \etaT k^{2} = (\Ca-1) \etaT k^{2},
\EN
where $\Ca=\alpha/(\etaT k)$ is the relevant dynamo number for the
$\alpha^2$ dynamo, $\alpha$ is the $\alpha$ coefficient
which is proportional to the small-scale kinetic helicity,
and $\etaT = \eta + \eta_{\rm t}$ is the
sum of molecular and turbulent magnetic diffusivity.
Clearly, dynamo action occurs when $|\Ca|>\Cacrit$, where the onset
condition is $\Cacrit=1$.
Standard estimates for isotropic turbulence in the high conductivity limit
\cite{Radler1980book,MoffattBook1978} yield
$\alpha\approx-(\tau/3)\bra{\oo\cdot\uu}$
and $\etat\approx(\tau/3)\bra{\uu^2}$, where $\tau$ is the correlation time
of the turbulence, $\oo=\nab\times\uu$ is the vorticity
and $\uu$ is the velocity in the small-scale fields.
Here, $\langle . \rangle$ denotes a volume average.
Using $\etat\gg\eta$, we have
\EQ
\Ca\approx-\bra{\oo\cdot\uu}/(k\bra{\uu^2}).
\EN
It is convenient to define $\bra{\oo\cdot\uu}/(\kf\bra{\uu^2})$ as the
normalized kinetic helicity, $\epsf$, so $\Ca\approx-\epsf\kf/k$.
This scaling implies that the critical value of the normalized helicity
$\epsf$ scales inversely proportional to the scale separation ratio,
i.e.\ $\epsf^{\rm crit} \propto(\kf/k)^{-1}$, where $k\ll\kf$ is the wave number
of the resulting large-scale magnetic field.
This wave number can be equal to $k=k_1\equiv2\pi/L$, which is the smallest
wave number in a periodic domain of size $L$.

In summary, the critical dynamo number $\Cacrit$, which decides between
growing or decaying solutions of the large-scale dynamo (LSD),
is proportional to the product of normalized helicity $\epsf$
and scale separation ratio $\kf/k$.
Therefore, the amount of helicity needed for the LSD is inversely proportional
to the scale separation ratio, and not some higher power of it.
It should be noted that the {\em normalized} kinetic helicity
$\epsf$ used here is not the same as the {\em relative} kinetic helicity,
$\tilde\epsf=\bra{\oo\cdot\uu}/(\orms\urms)$.
The two are related to each other via the relation
\EQ
\tilde\epsf/\epsf=(\kom/\kf)^{-1},
\EN
where $\kom\approx\orms/\urms$ is inversely proportional to the
Taylor microscale.
Here, the subscripts rms refer to root-mean-square values.
For small Reynolds numbers, $\kom$ provides a useful
estimate of the wave number $\kf$ of the energy-carrying eddies.
In contrast, for large Reynolds numbers $\Rey$, we expect $\kom/\kf$ to be
proportional to $\Rey^{1/2}$, so $\tilde\epsf$ decreases correspondingly
while $\epsf$ remains unchanged.

To understand the saturation of a helical dynamo, it is important to
understand the relation between the resulting large-scale field and
the associated small-scale field.
Indeed, the growth of the large-scale field is always accompanied
by a growth of small-scale magnetic field.
Small-scale here means the scale of the underlying turbulent
motions, which drive the dynamo.
Conservation of total magnetic helicity causes a build-up of magnetic helicity
at large scales and of opposite sign at small scales
\cite{Seehafer1996,Ji1999}.
As the dynamo saturates, the largest scales of the magnetic field become even
larger, which finally leads to a field of a scale that is similar to that
of the system itself.
This can be understood as being the result of an inverse cascade,
which was first predicted based on closure calculations
\cite{Frisch-Pouquet-Leorat-1975-JFluidMech}.

If the domain is closed or periodic,
the build-up of small-scale magnetic helicity causes the
$\alpha$ effect to diminish, which marks the end of the exponential growth
and could occur well before final saturation is reached.
The dynamo then is said to be catastrophically quenched and,
in a closed or periodic system, the subsequent growth to the final state
happens not on a dynamical timescale, but on a resistive one.
Quenching becomes stronger as the magnetic Reynolds number increases, which,
for astrophysically relevant problems, means a total loss of the LSD
within the timescales of interest.
In the case of open boundaries magnetic helicity fluxes can occur, which
can alleviate the quenching and allow for fast saturation of the large-scale
magnetic field
\cite{BlackmanField2000a,BlackmanField2000b,Kleeorin_etal2000,ssHelLoss09}.

In a recent publication \cite{GrahamBlackmanMinuteHelicity12}
it was argued that for periodic boundaries
the critical value of $\epsf$ for LSD action to occur
decreases with the scale separation ratio like
$\epsf^{\rm crit} \propto (\kf/k_1)^{-3}$.
Their finding, however, is at variance with the predictions made
using equation \eqref{eq: lambda kinetic}, which would rather suggest
a dependence of $\epsf^{\rm crit} \propto (\kf/k_1)^{-1}$ with $\Cacrit = 1$.
This discrepancy could be a consequence of the criterion used
in \cite{GrahamBlackmanMinuteHelicity12} for determining $\Cacrit$.
The authors looked at the growth rate of the magnetic field after the
end of the kinematic growth phase, but only at a small fraction of the
resistive time.
Therefore their results might well be contaminated by
magnetic fields resulting from the small-scale dynamo (SSD).
Earlier simulations \cite{Brandenburg2009ApJ} have demonstrated that for
$\Rm\ge100$, the growth rate of the helical LSD approaches the well-known
scaling of the nonhelical SSD with $\lambda\propto\Rey^{1/2}$,
which corresponds to the turnover rate of the smallest turbulent eddies
\cite{Schekochihin_etal2004ApJ,Haugen_etal2004PhRvE}.

Given that the LSD is best seen in the nonlinear regime
\cite{BrandInverseCascade2001}, we decided to determine $\Cacrit$
from a bifurcation diagram by extrapolating to zero.
In a bifurcation diagram, we plot the energy of the mean or
large-scale field versus $\Ca$.
Simple considerations using the magnetic helicity equation applied
to a homogeneous system in the steady state show that the current
helicity must vanish \cite{BrandInverseCascade2001}.
In a helically driven system, this implies that the current helicity of
the large-scale field must then be equal to minus the current helicity
of the small-scale field.
For a helical magnetic field, the normalized mean square magnetic
field, $\bra{\meanBB^2}/\Beq^2$, is approximately equal to $\Ca-\Cacrit$.
Here, $\Beq = (\mu_{0}\mean{\rho})^{1/2}\urms$ is the equipartition value
of the magnetic field, $\mu_{0}$ is the vacuum permeability,
and $\mean{\rho}$ is the mean density.
Again, since $\Cacrit\approx1$ and $\Ca\approx\epsf\kf/k_1$,
this suggests that the LSD is excited for $\epsf>(\kf/k_1)^{-1}$
rather than some higher power of $\kf/k_1$.
This is a basic prediction that has been obtained from nonlinear
mean-field dynamo models that incorporate magnetic helicity evolution
\cite{BlackmanBrandenb2002ApJ} as well as from direct numerical
simulations in the presence of shear \cite{KapylaBrandenb2009ApJ}.
It is important to emphasize that mean field dynamo theory has been
criticized on the grounds that no $\alpha$ effect may exist in the
highly nonlinear regime at large magnetic Reynolds numbers \cite{CH09}.
This is however in conflict with results of numerical simulations using
the test-field method \cite{BRRS08} showing that $\alpha$ effect and
turbulent diffusivity are both large, and that only the difference
between both effects is resistively small.
Another possibility is that the usual helical dynamo of $\alpha^2$ type
may not be the fastest growing one \citep{BCR05}.
This is related to the fact that, within the framework of the
Kazantsev model \cite{Kaz68} with helicity, there are new solutions
with long-range correlations \citep{S99,BS00}, which
could dominate the growth of a large scale field at early times.
The purpose of the present paper is therefore to reinvestigate the
behavior of solutions in the nonlinear regime over
a broader parameter range in the light of recent conflicting findings
\cite{GrahamBlackmanMinuteHelicity12}.

\section{The model}

\subsection{Basic equations}

Following earlier work, we solve the compressible hydromagnetic equations
using an isothermal equation of state.
Although compressibility is not crucial for the present purpose, it does
have the advantage of avoiding the nonlocality associated with solving
for the pressure, which requires global communication.
Thus, we solve the equations
\begin{eqnarray}
&& \frac{\partial}{\partial t} \AAA = \UU\times\BB -\eta\mu_{0}\JJ,
\label{eq: induction} \\
&& \frac{\DD}{\DD t} \UU = -\cs^{2}\nab \ln{\rho} +
\frac{1}{\rho} \JJ\times\BB + \BoldVec{F}_{\rm visc} + \BoldVec{f},
\label{eq: momentum} \\
&& \frac{\DD}{\DD t} \ln{\rho} = -\nab\cdot\UU, \label{eq: continuity}
\end{eqnarray}
where $\AAA$ is the magnetic vector potential, $\UU$ the velocity,
$\BB$ the magnetic field, $\eta$ the molecular magnetic diffusivity,
$\mu_{0}$ the vacuum permeability, $\JJ$ the electric
current density, $\cs$ the isothermal sound speed, $\rho$ the
density, $\BoldVec{F}_{\rm visc}$ the viscous force,
$\BoldVec{f}$ the helical forcing term,
and $\DD/\DD t = \partial/\partial t + \UU\cdot\nab$ the advective time
derivative.
The viscous force is given as
$\FF_{\rm visc} = \rho^{-1}\nab\cdot2\nu\rho\SSSS$,
where $\nu$ is the kinematic viscosity,
and $\SSSS$ is the traceless rate of strain tensor with components
${\sf S}_{ij}=\frac{1}{2}(u_{i,j}+u_{j,i})-\frac{1}{3}\delta_{ij}\nab\cdot\UU$.
Commas denote partial derivatives.

The energy supply for a helically driven dynamo is provided by the
forcing function $\BoldVec{f} = \BoldVec{f}(\xx,t)$, which is a helical
function that is random in time.
It is defined as
\EQ
\ff(\xx,t)={\rm Re}\{N\ff_{\kk(t)}\exp[\ii\kk(t)\cdot\xx+\ii\phi(t)]\},
\label{ForcingFunction}
\EN
where $\xx$ is the position vector.
The wave vector $\kk(t)$ and the random phase
$-\pi<\phi(t)\le\pi$ change at every time step, so $\ff(\xx,t)$ is
$\delta$-correlated in time.
For the time-integrated forcing function to be independent
of the length of the time step $\delta t$, the normalization factor $N$
has to be proportional to $\delta t^{-1/2}$.
On dimensional grounds it is chosen to be
$N=f_0 c_{\rm s}(|\kk|c_{\rm s}/\delta t)^{1/2}$, where $f_0$ is a
nondimensional forcing amplitude.
We choose $f_0=0.02$, which results in a maximum Mach number of about 0.3
and an rms value of about 0.085.
At each timestep we select randomly one of many possible wave vectors
in a certain range around a given forcing wave number.
The average wave number is referred to as $k_{\rm f}$.
Transverse helical waves are produced via \cite{Haugen_etal2004PhRvE}
\begin{equation}
\ff_{\kk}=\RRRR\cdot\ff_{\kk}^{\rm(nohel)}\quad\mbox{with}\quad
{\sf R}_{ij}={\delta_{ij}-\ii\sigma\epsilon_{ijk}\hat{k}_k
\over\sqrt{1+\sigma^2}},
\label{eq: forcing}
\end{equation}
where $\sigma$ is a measure of the helicity of the forcing and
$\sigma=1$ for positive maximum helicity of the forcing function.
Furthermore,
\EQ
\ff_{\kk}^{\rm(nohel)}=
\left(\kk\times\eee\right)/\sqrt{\kk^2-(\kk\cdot\eee)^2}
\label{nohel_forcing}
\EN
is a nonhelical forcing function, where $\eee$ is an arbitrary unit vector
not aligned with $\kk$; note that $|\ff_{\kk}|^2=1$ and
\EQ
\ff_{\kk}\cdot(\ii\kk\times\ff_{\kk})^*=2\sigma k/(1+\sigma^2),
\EN
so the relative helicity of the forcing function in real space is
$2\sigma/(1+\sigma^2)$.

For comparison with earlier work, we shall also use in one case
an ABC-flow forcing function \citep{Gal12},
\EQ
\ff(\xx)={f_0\over\sqrt{{3\over2}(1+\sigma^2)}}\left(
\begin{matrix}
\sin X_3+\sigma\cos X_2\\
\sin X_1+\sigma\cos X_3 \\
\sin X_2+\sigma\cos X_1 
\end{matrix}
\right),
\label{ABC}
\EN
where $X_i=\kf x_i+\theta_i$ and $\theta_i=\theta_0\cos\omega_i t$
are time-dependent phases that vary sinusoidally with frequencies $\omega_i$
and amplitude $\theta_0$.
This forcing function is easy to implement and
serves therefore as a proxy of helical turbulence; see
Refs.~\cite{Galanti_etal1992GApFD,GrahamBlackmanMinuteHelicity12},
where the phases changed randomly.
We have restricted ourselves to the special case where the coefficients
in front of the trigonometric functions are unity, but those could be
made time-dependent too; see Ref.~\cite{Kleeorin_etal2009PhRvE}.
However, as we will see below, ABC-flow driven dynamos do not
show some crucial aspects of random plane wave-forced helical turbulence.
Most of the results presented below concern the forcing function
\Eq{ForcingFunction}, and only one case with \Eq{ABC} will be
considered at the end.

Our model is governed by several nondimensional parameters.
In addition to the scale separation ratio $\kf/k_1$, introduced above,
there are the magnetic Reynolds and Prandtl numbers
\EQ
\Rm=\urms/(\eta\kf),\quad
\Pm=\nu/\eta.
\label{Rey_def}
\EN
These two numbers also define the fluid Reynolds number,
$\Rey=\urms/(\nu\kf)=\Rm/\Pm$.
The maximum values that can be attained are limited by the numerical
resolution and become more restrictive at larger scale separation.
The calculations have been performed using the {\sc Pencil Code} (see
http://pencil-code.googlecode.com) at resolutions of up to $512^3$
mesh points.

\subsection{Mean-field interpretation}

The induced small-scale motions $\uu$ are helical and give rise to the
usual (kinetic) $\alpha$ effect \cite{Radler1980book}
\EQ
\alpha_{\rm K} \approx -\frac{\langle\oo\cdot\uu\rangle}{3\urms\kf}.
\label{eq: alphaK}
\EN
In the nonlinear regime, following the early work of Pouquet, Frisch,
and L\'eorat \cite{PouquetFrischLeorat1976JFM}, the relevant $\alpha$
effect for dynamo action is believed to be the sum of the kinetic and
a magnetic $\alpha$, i.e.,
\EQ
\alpha \approx \frac{-\langle\oo\cdot\uu\rangle +
\langle\jj\cdot\bb\rangle/\bra{\rho}}{3\urms\kf}.
\label{eq: alphaK+alphaM}
\EN
Simulations have confirmed the basic form of \Eq{eq: alphaK+alphaM}
with equal contributions from $\langle\oo\cdot\uu\rangle$ and
$\langle\jj\cdot\bb\rangle/\bra{\rho}$, but one may argue that the
second term should only exist in the presence of hydromagnetic background
turbulence \cite{RR07}, and not if the magnetic fluctuations are a
consequence of tangling of a mean field produced by dynamo action
as in the simulations in Ref.~\cite{BrandInverseCascade2001}.
However, to explain the resistively slow saturation
in those simulations, the only successful explanation
\cite{FieldBlackman2002ApJ,BlackmanBrandenb2002ApJ} comes from
considering the magnetic helicity equation, which feeds back
onto the $\alpha$ effect via \Eq{eq: alphaK+alphaM}.
This is our main argument in support of the applicability of
this equation.
Another problem with \Eq{eq: alphaK+alphaM} is the assumption of isotropy
\cite{RR07}, which has however been relaxed in subsequent work \cite{BS07}.
Let us also mention that \Eq{eq: alphaK+alphaM} is usually obtained using
the $\tau$ approximation.
In its simplest form, it yields incorrect results in the low conductivity
limit, where the second-order correlation approximation applies
\cite{Radler1980book,MoffattBook1978}.
However, this is just a consequence of making simplifying assumptions
in handling the diffusion operator, which can be avoided, too \cite{SSB07}.
At higher conductivity, numerical simulations have been able to reproduce
some important predictions from the $\tau$ approximation \cite{BKM04}.

\EEq{eq: alphaK+alphaM} is used to derive the expression for the
resistively slow saturation behavior \cite{FieldBlackman2002ApJ}.
We will not reproduce here the derivation, which can be found
elsewhere \cite{BlackmanBrandenb2002ApJ}.
The resulting large-scale fields can be partially helical,
which means one can write
\EQ
\langle\JJ\cdot\BB\rangle = \epsilon_{\rm m} k_{\rm m} \langle\BB^{2}\rangle,
\label{eq: epsmdef}
\EN
with large-scale wave vector $k_{\rm m}$ and corresponding fractional helicity
$\epsilon_{\rm m}$, defined through \Eq{eq: epsmdef}.
However, in the cases considered below the domain is triply periodic,
so the solutions are Beltrami fields for which
$k_{\rm m}\approx k_1$ and $\epsilon_{\rm m}\approx1$ is an
excellent approximation, and only $\epsf$ will take values less unity.
Nevertheless, in some expressions we retain the $\epsm$ factor for clarity.
For example, the saturation value of the large-scale magnetic field,
$B_{\rm sat}$, is given by \cite{BlackmanBrandenb2002ApJ}
\EQ
B_{\rm sat}^2/\Beq^2 \approx \left(|C_{\alpha}|/\epsm - 1\right)\iota,
\label{eq: Bsat MF}
\EN
where $\Ca=\alpha_{\rm K}/(\etaT k_1)$ is the relevant dynamo number
based on the smallest wavenumber in the domain and
$\iota=1+3/\Rm\equiv\etaT/\etat$ is a correction factor resulting from
the fact that $\etaT$ is slightly bigger than $\etat$.
The factor $3$ in the expression for $\iota$ results from our definition
of $\Rm$ and the fact that \cite{Sur_etal2008}
\EQ
\etat\approx\urms/(3\kf)=\eta\Rm/3.
\label{eq: etat}
\EN
\EEq{eq: Bsat MF} shows clearly the onset condition
$|\Ca| > |\epsilon_{\rm m}|\approx1$.
Using \Eqs{eq: alphaK}{eq: etat}, we find
\EQ
\Ca\approx-{\bra{\oo\cdot\uu}\over\iota k_1\urms^2}=-{\epsf\kf\over\iota k_1}.
\EN
From \Eq{eq: Bsat MF} we can derive the critical value of the
normalized helicity $\epsf$ as a function of the scale separation ratio.
Setting $\Ca$ to its critical value ($|\Ca| = \epsilon_{\rm m}$) we obtain
\EQ
\epsfcrit\approx\iota\epsm\left(\frac{\kf}{k_1}\right)^{-1},
\label{eq: epsf crit}
\EN
which is at variance with the findings in
Ref.~\cite{GrahamBlackmanMinuteHelicity12}.

Once the dynamo is excited and has reached a steady state,
not only $\alpha$ but also $\etat$ will be suppressed.
This can be taken into account using a quenching factor
$g(\meanB)$, so $\etat(\meanB)=\etatz g(\meanB)$ with
$g=(1+\tilde{g}|\meanBB|/\Beq)^{-1}$ \cite{KRP94,RK01,BrandInverseCascade2001}.
\EEq{eq: Bsat MF} is then modified and reads
$B_{\rm sat}^2/\Beq^2=(|C_{\alpha}|-C_{\alpha0})\iota/\epsm$
with 
\EQ
C_{\alpha0}=[1-(1-g)/\iota]\epsm.
\label{eq: Bsat MF mod}
\EN
Note that $C_{\alpha0}=\epsm^{-1}$ in the unquenched case,
i.e., for $g=1$.

\subsection{Simulation strategy}
\label{Strategy}

We recall that our forcing term $\BoldVec{f}$
in equation \Eq{ForcingFunction}
is a stochastic forcing centered around the wave number $\kf$.
In contrast to Ref.~\cite{GrahamBlackmanMinuteHelicity12}, this forcing is
$\delta$-correlated in time.
The fractional helicity of the helical forcing is a free parameter.
The simulation domain is a periodic cube with dimensions $2\pi$.
Due to the cubic geometry of the domain, the large-scale magnetic field
can orient itself in three possible directions.
Therefore, we compute three possible planar averages
($xy$, $xz$, and $yz$ averages).
From their resistive evolution we infer their saturation values 
at the end of the resistive phase.
The strongest field gives then the relevant mean-field $\meanBB$.

\begin{figure}[t!]\begin{center}
\includegraphics[width=\columnwidth]{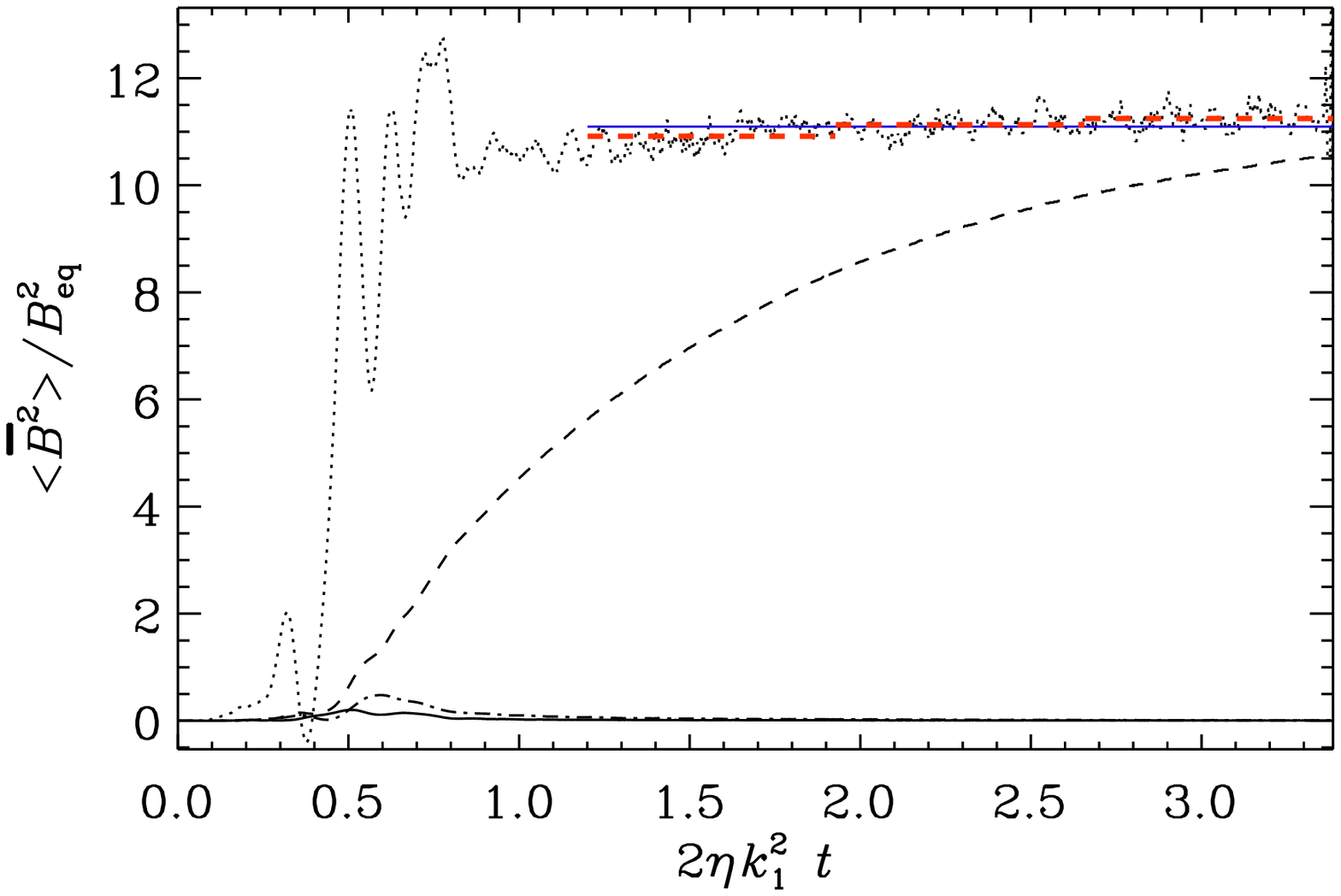}
\end{center}\caption[]{
(Color online)
Example showing the evolution of the normalized $\bra{\meanBB^2}$ (dashed)
and that of $\bra{\meanBB^2}+\tau\dd\bra{\meanBB^2}/\dd t$ (dotted),
compared with its average in the interval $1.2\leq2\eta k_1^2t\leq3.5$
(horizontal blue solid line), as well as averages over three subintervals
(horizontal red dashed lines).
Here, $\meanBB$ is evaluated as an $xz$ average, $\bra{\BB}_{xz}$.
For comparison we also show the other two averages, $\bra{\BB}_{xy}$
(solid) and $\bra{\BB}_{yz}$ (dash-dotted), but their values are very small.
}\label{psat}\end{figure}

Since $\meanBB$ is helical and magnetic helicity can only change
on resistive timescales, the temporal evolution of the energy of the
mean magnetic field, $M(t)$, is given by \cite{BrandInverseCascade2001}
\EQ
M(t)=M_0-M_1 e^{-t/\tau},
\label{eq: sat}
\EN
where $\tau^{-1}=2\eta\epsilon_{\rm m}^2 k_1^2$ is known,
$M_0=B_{\rm sat}^2$ is the square of the desired saturation field strength,
and $M_1$ is an unknown constant that can be positive or negative,
depending on whether the initial magnetic field of a given calculation
was smaller or larger than the final value.
(Here, an initial field could refer to the last snapshot of another
calculation with similar parameters, for example.)
The functional behavior given by \Eq{eq: sat}
allows us to determine $B_{\rm sat}^2$ as the time average of
$M+\tau\dd M/\dd t$, which should only fluctuate about a constant value, i.e.,
\EQ
B_{\rm sat}^2\approx{1\over t_2-t_1}\int_{t_1}^{t_2}
\left[\bra{\meanBB^2}(t')+\tau{\dd\over\dd t'}\bra{\meanBB^2}\right]\,\dd t'.
\EN
This technique has the advantage that we do not need to wait until
the field reaches its final saturation field strength.
Error bars can be estimated by computing this average for each third
of the full time series and taking the largest departure from the
average over the full time series.
An example is shown in \Fig{psat}, where we see $\bra{\meanBB^2}$
still growing while  $\bra{\meanBB^2}+\tau\dd\bra{\meanBB^2}/\dd t$
is nearly constant when $\bra{\meanBB^2}$ reaches a value less
than half its final one.
This figure shows that the growth of $\bra{\meanBB^2}$ follows the
theoretical expectation \Eq{eq: sat} quite closely and that temporal
fluctuations about this value are small, as can be seen by the fact
that its time derivative fluctuates only little. 

\section{Results}

\subsection{Dependence of kinetic helicity on $\sigma$}

\begin{figure}[t!]\begin{center}
\includegraphics[width=\columnwidth]{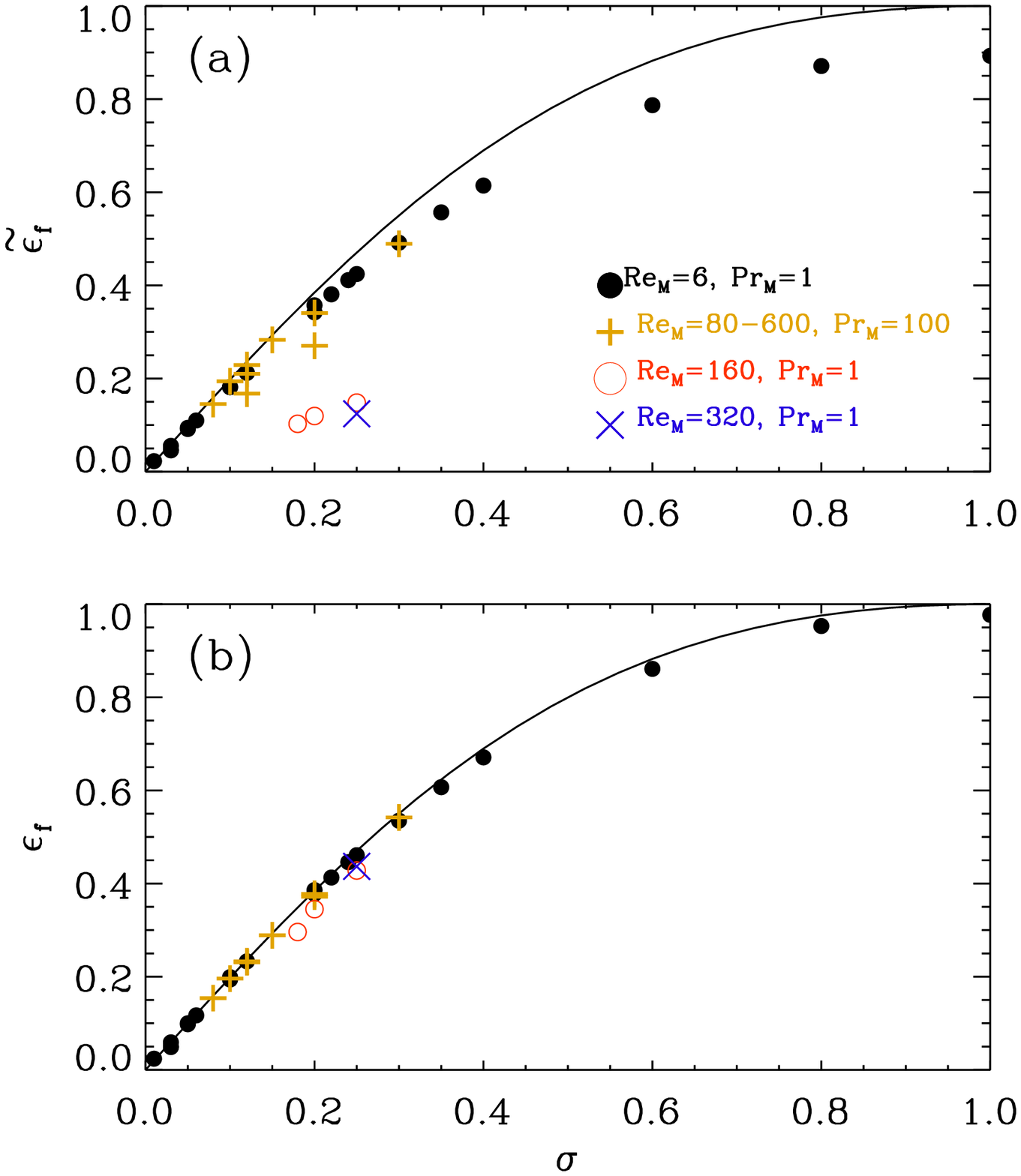}
\end{center}\caption[]{
(Color online)
Dependence of relative kinetic helicity $\tilde\epsf$ (a) and normalized
kinetic helicity $\epsf$ (b) on the helicity parameter $\sigma$ of the
forcing function \Eq{eq: forcing} together with the analytical expression
$2\sigma/(1+\sigma^{2})$ (solid line).
}\label{presults_epsf}\end{figure}

We recall that the relative helicity of the forcing function is
$\bra{\ff\cdot\nab\times\ff}/[\ff_{\rm rms}(\nab\times\ff)_{\rm rms}]
=2\sigma/(1+\sigma^2)$.
This imposes then a similar variation onto the relative kinetic helicity,
$\tilde\epsf=\bra{\oo\cdot\uu}/(\orms\urms)$; see \Fig{presults_epsf}(a).
However, as discussed above, $\tilde\epsf$ is smaller than $\epsf$
by a factor $\kom/\kf$, which in turn depends on the Reynolds number (see below).
It turns out that $\epsf$ matches almost exactly the values of
$2\sigma/(1+\sigma^2)$; see \Fig{presults_epsf}(b).

\begin{figure}[t!]\begin{center}
\includegraphics[width=\columnwidth]{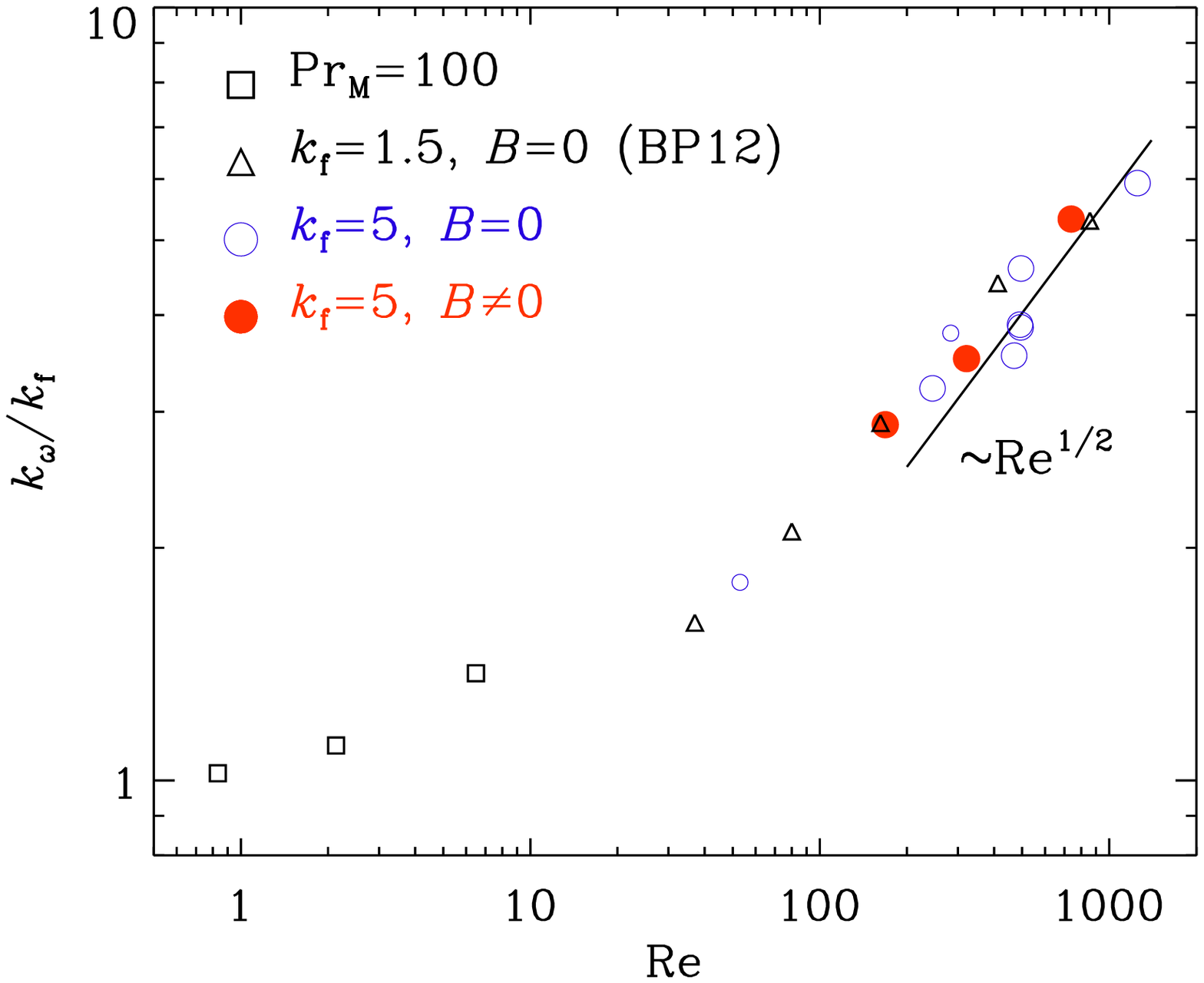}
\end{center}\caption[]{
(Color online)
Dependence of $\kom/\kf$ on $\Rey$.
The open and closed circles correspond to runs with $\Pm=1$ without and
with magnetic field, respectively, while squares correspond to runs with
$\Pm=100$ and $\Rey=\Rm/\Pm$ is small.
Triangles denote the results for $\kf/k_1=1.5$ of Ref.~\cite{BP12} (BP12).
}\label{pkom}\end{figure}

The theoretically expected scaling $\kom/\kf\propto\Rey^{1/2}$
is a well-known result for high Reynolds number turbulence \cite{Batchelor1953},
and has recently been verified using simulations similar to those
presented here, but without magnetic field and a smaller scale separation
ratio of $\kf/k_1=1.5$ \citep{BP12}.
For our current data we find that such a scaling is obeyed for $\Pm=1$
and large values of $\Rey$, independently of the presence of magnetic
field or kinetic helicity, but this scaling is not obeyed when $\Pm=100$
and $\Rey$ is small; see \Fig{pkom}.

\subsection{Dependence on scale separation}

Next, we perform simulations with different forcing wave numbers $\kf$
and different values of $\epsf$ at
approximately constant magnetic Reynolds number,
$\Rm \approx 6$, and fixed magnetic Prandtl number, $\Pm = 1$.
Near the end of the resistive saturation phase we look at the energy of
the strongest mode at $k=k_1$, using the method described in \Sec{Strategy}.
We choose this rather small value of $\Rey$ because we want to access
relatively large scale separation ratios of up to $\kf/k_1=80$.
Given that the Reynolds number based on the scale of the domain
is limited by the number of mesh points (500, say), it follows that
for $\kf/k_1=80$ the Reynolds number defined through \Eq{Rey_def} is 6.
For comparison, a Reynolds number based on the size of the domain,
i.e., $\urms L/\eta$, would be larger by a factor $2\pi$, i.e., 3000.

\begin{figure}
\begin{center}
\includegraphics[width=\columnwidth]{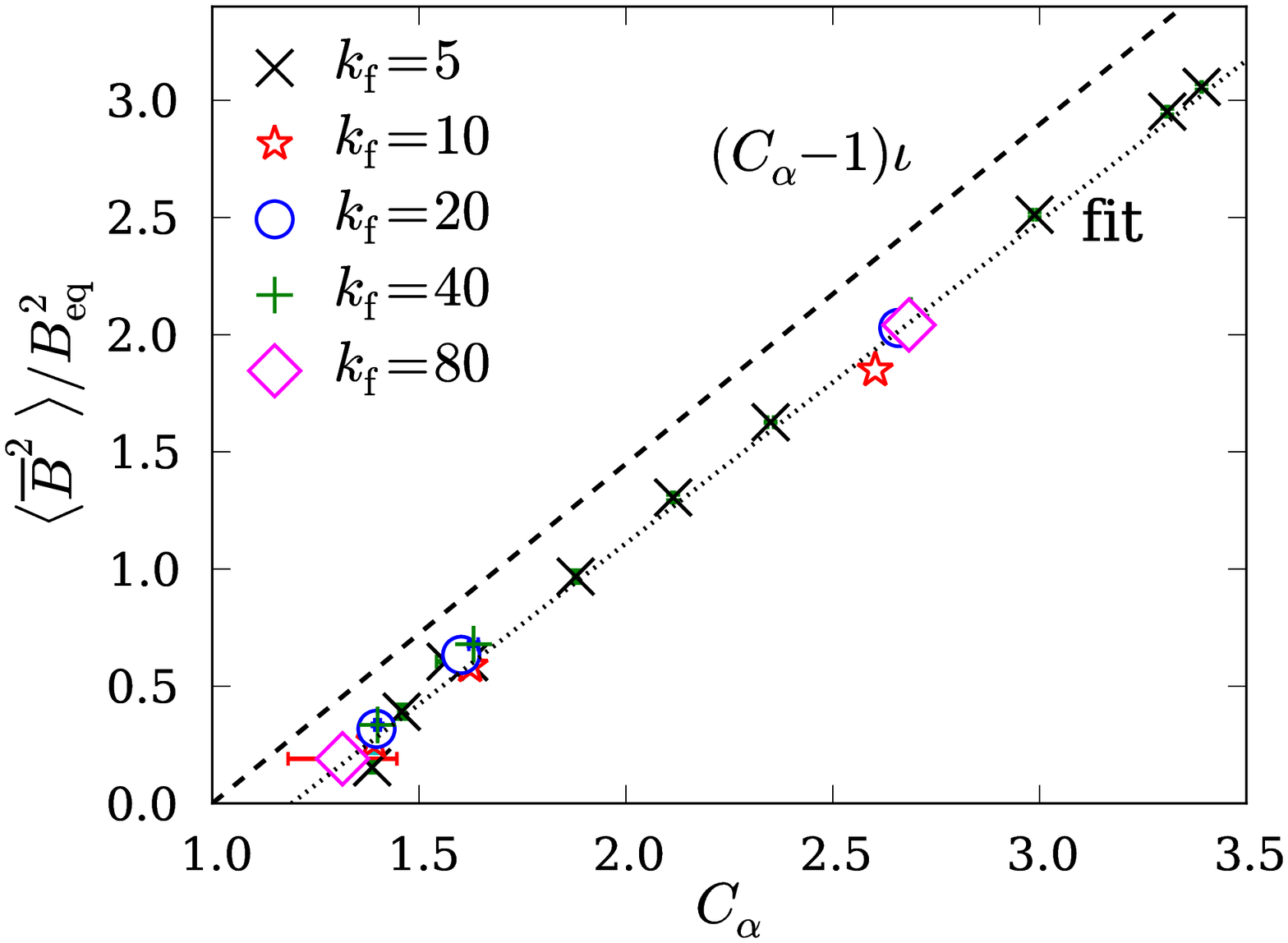}
\end{center}\caption[]{
(Color online)
Steady state values of $\bra{\meanBB^2}/\Beq^2$ as a function of
$C_{\alpha}$ together with the theoretical prediction
from \Eq{eq: Bsat MF} (dashed line) and a linear fit (dotted line).
}\label{fig: Bsat_Ca}
\end{figure}

\begin{figure}
\begin{center}
\includegraphics[width=\columnwidth]{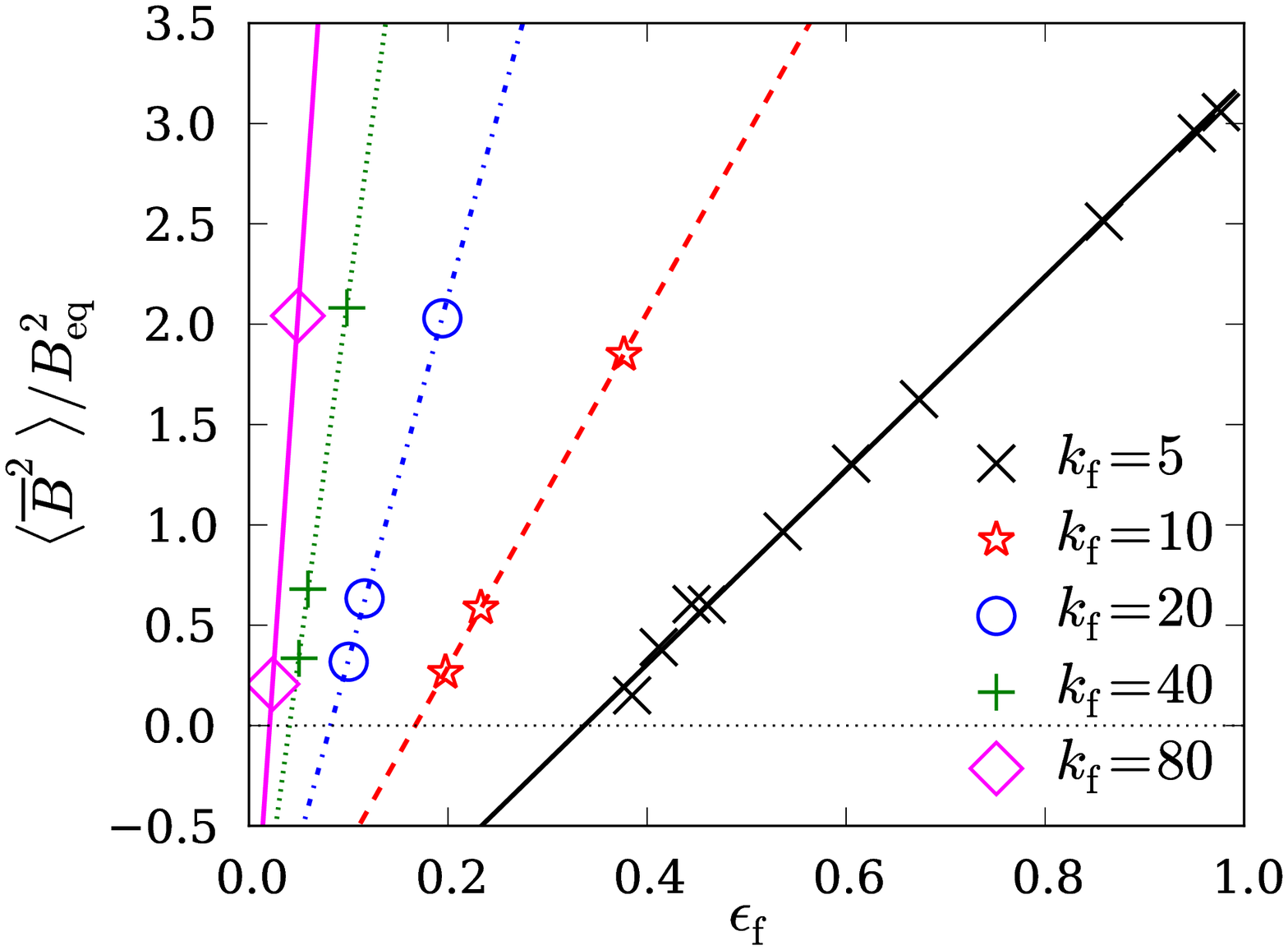}
\end{center}\caption[]{
(Color online)
Steady state values of $\bra{\meanBB^2}/\Beq^2$ as a function of
$\epsf$ for various scale
separation values $\kf/k_1$ together with linear fits.
}\label{fig: ef_crit_fits}
\end{figure}

\begin{figure}
\begin{center}
\includegraphics[width=\columnwidth]{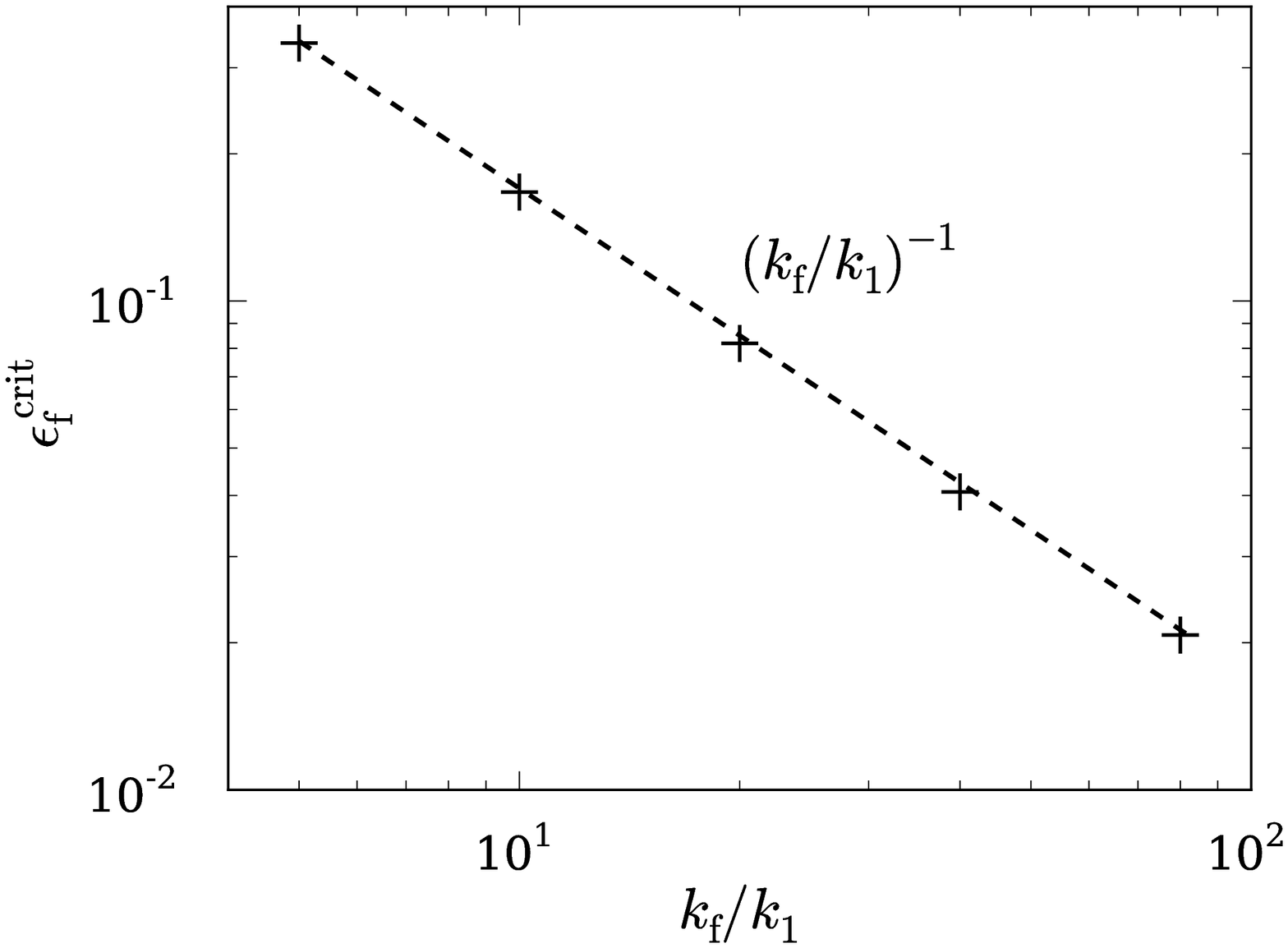}
\end{center}\caption[]{
Critical value for the normalized kinetic helicity $\epsf$
for which LSD action occurs for different scale separations.
}\label{fig: ef_crit_kf}
\end{figure}

As seen from \Eq{eq: Bsat MF}, mean-field considerations predict
a linear increase of the saturation magnetic energy with
$\Ca$ and onset at $\Ca = 1$.
This behavior is reproduced in our simulation (\Fig{fig: Bsat_Ca}), where
we compare the theoretical prediction with the simulation results.
For different values of $\kf/k_1$ and $\Ca$ we extrapolate
the critical value $\Cacrit\approx1.2$ (\Fig{fig: Bsat_Ca}),
which gives the critical values
$\epsfcrit\approx1.2\iota\,(\kf/k_1)^{-1}=1.7\,(\kf/k_1)^{-1}$ for which
the LSD is excited.
For each scale separation value we plot the dependence of
$\bra{\meanBB^2}/\Beq^2$ on $\epsf$ (\Fig{fig: ef_crit_fits}) and make linear
fits.
From these fits we can extrapolate the critical values $\epsfcrit$, for which
the LSD gets excited (\Fig{fig: ef_crit_kf}), which gives again
$\epsfcrit\approx1.7\,(\kf/k_1)^{-1}$.

\begin{figure}
\begin{center}
\includegraphics[width=\columnwidth]{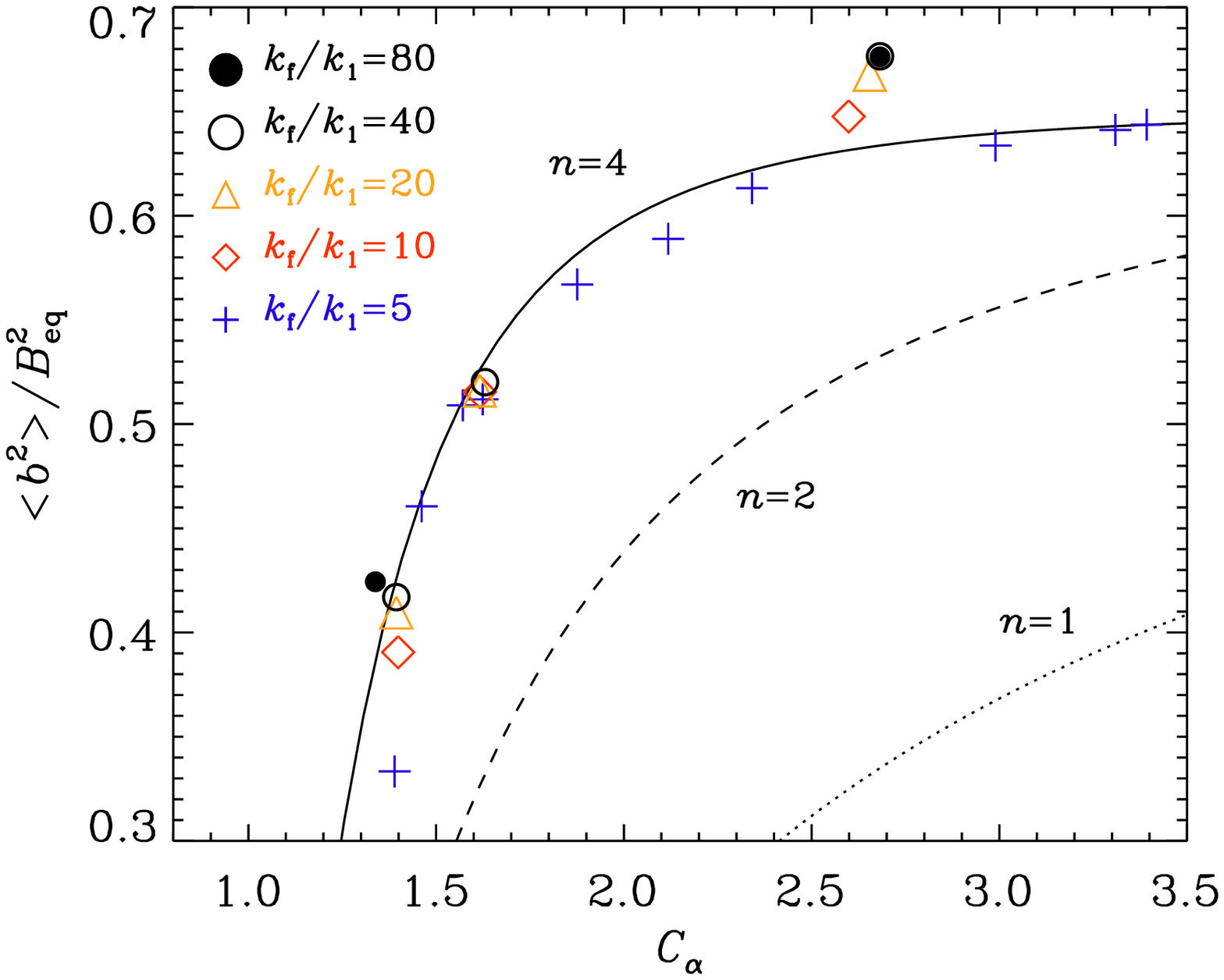}
\end{center}\caption[]{
(Color online)
Steady state values of $\bra{\bb^2}/\Beq^2$ as a function of
$C_{\alpha}$ together with the fit formula from \Eq{eq: b2fit}
with $n=4$, compared with $n=1$ (dotted) and $n=2$ (dashed).
Different symbols denote different values of $\kf/k_1$.
}\label{fig: presults_flucts}
\end{figure}

It is noteworthy that the graph of $\bra{\meanBB^2}/\Beq^2$ versus $\Ca$
deviates systematically (although only by a small amount)
from the theoretically expected value, $(\Ca-1)\iota$.
While the slope is rather close to the expected one, the LSD onset
is slightly delayed and occurs at $\Ca\approx1.2$
instead of 1.
The reason for this is not clear, although one might speculate that
it could be modeled by adopting modified effective values of $\iota$
or $\epsm$ in \Eq{eq: Bsat MF mod}.
Apart from such minor discrepancies with respect to the simple theory,
the agreement is quite remarkable.
Nevertheless, we must ask ourselves whether this agreement persists for
larger values of the magnetic Reynolds number.
This will be addressed in \Sec{DependenceOnRm}.

At this point we should note that there
is also a theoretical prediction for the energy in the magnetic
fluctuations, namely $\bra{\bb^2}/\Beq^2\approx(\Ca-\Cacrit)/\Ca$.
Nonetheless, the results shown in \Fig{fig: presults_flucts} deviate
from this relation and are better described by a modified formula
\EQ \label{eq: b2fit}
\bra{\bb^2}/\Beq^2\propto 1-({\Cacrit/\Ca})^n
\quad\mbox{(with $n\approx4$)}.
\EN
Again, the reason for this departure is currently unclear.

\subsection{Dependence on $\Rm$}
\label{DependenceOnRm}

To examine whether there is any unexpected dependence of the onset
and the energy of the mean magnetic field on $\Rm$ and to
approach the parameters used in Ref.~\cite{GrahamBlackmanMinuteHelicity12},
who used values up to $\Rm=1500$,
we now consider larger values of the magnetic Reynolds number.
This widens the inertial range significantly and leads to the
excitation of the SSD.
We consider first the case of a large magnetic Prandtl number
($\Pm=100$) and turn then to the more usual case of $\Pm=1$.
Our motivation behind the first case is that
higher values of $\Rm$ can more easily be reached at larger values of $\Pm$.
This is because at large values of $\Pm$, most of the
injected energy is dissipated viscously rather than resistively,
leaving less energy to be channeled down the magnetic cascade
\cite{Brandenburg2011AN}.
This is similar to the case of {\em small} values of $\Pm$,
where larger {\em fluid} Reynolds numbers can be reached because then most
of the energy is dissipated resistively \cite{Brandenburg2009ApJ}.
Here, however, we shall first be concerned with the former case of large
values of $\Pm$ and consider then the case of $\Pm=1$.

\begin{figure}
\begin{center}
\includegraphics[width=\columnwidth]{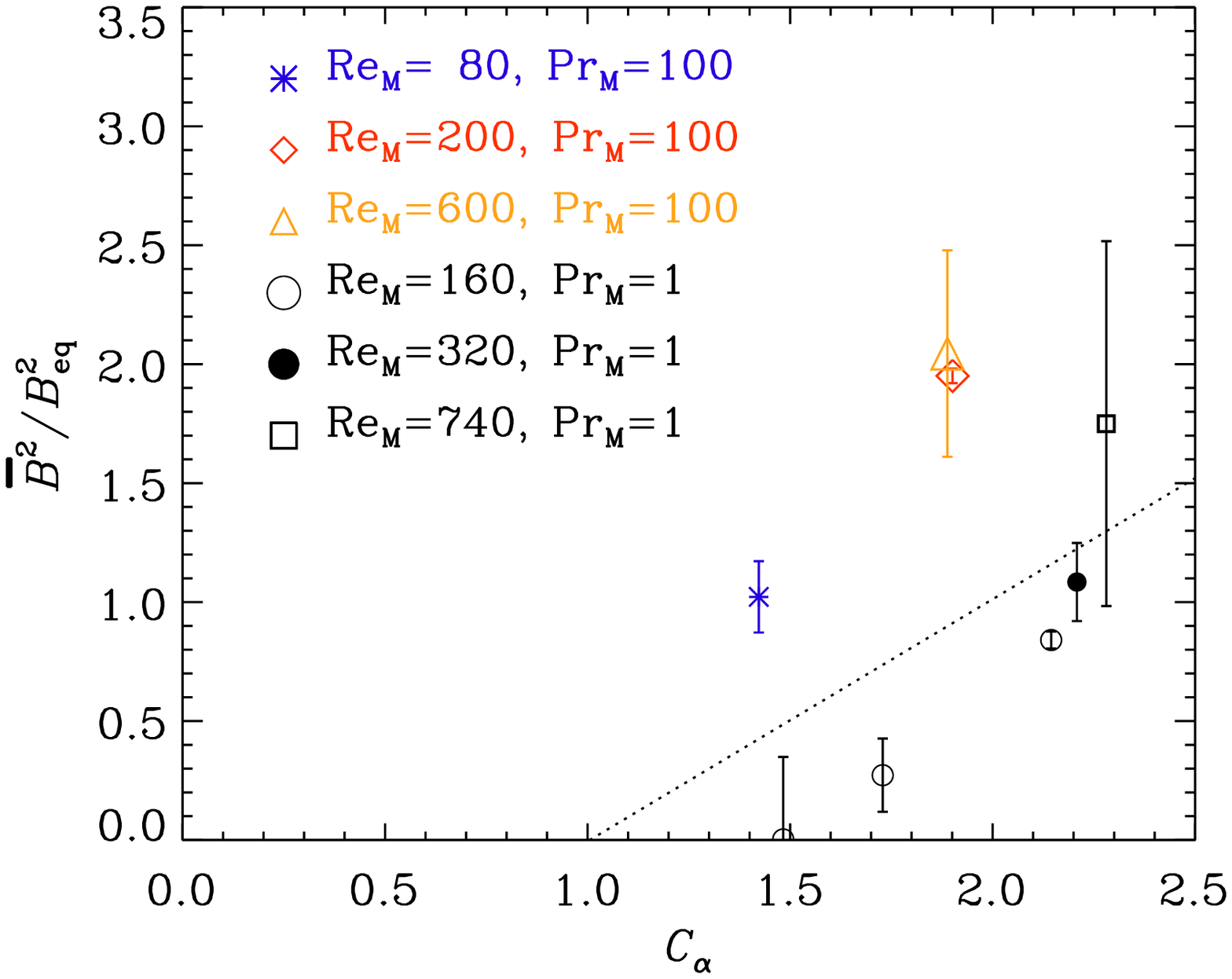}
\end{center}\caption[]{
(Color online)
Steady state values of $\bra{\meanBB^2}/\Beq^2$ as a function of
$C_{\alpha}$ for $\Pm=100$ and $\Pm=1$
for $\kf/k_1=5$ and different values of $\Rm$ (different symbols),
compared with the theoretical prediction (dotted line).
}\label{fig: Bsat_Ca_Pm100}
\end{figure}

\begin{figure}
\begin{center}
\includegraphics[width=.9\columnwidth]{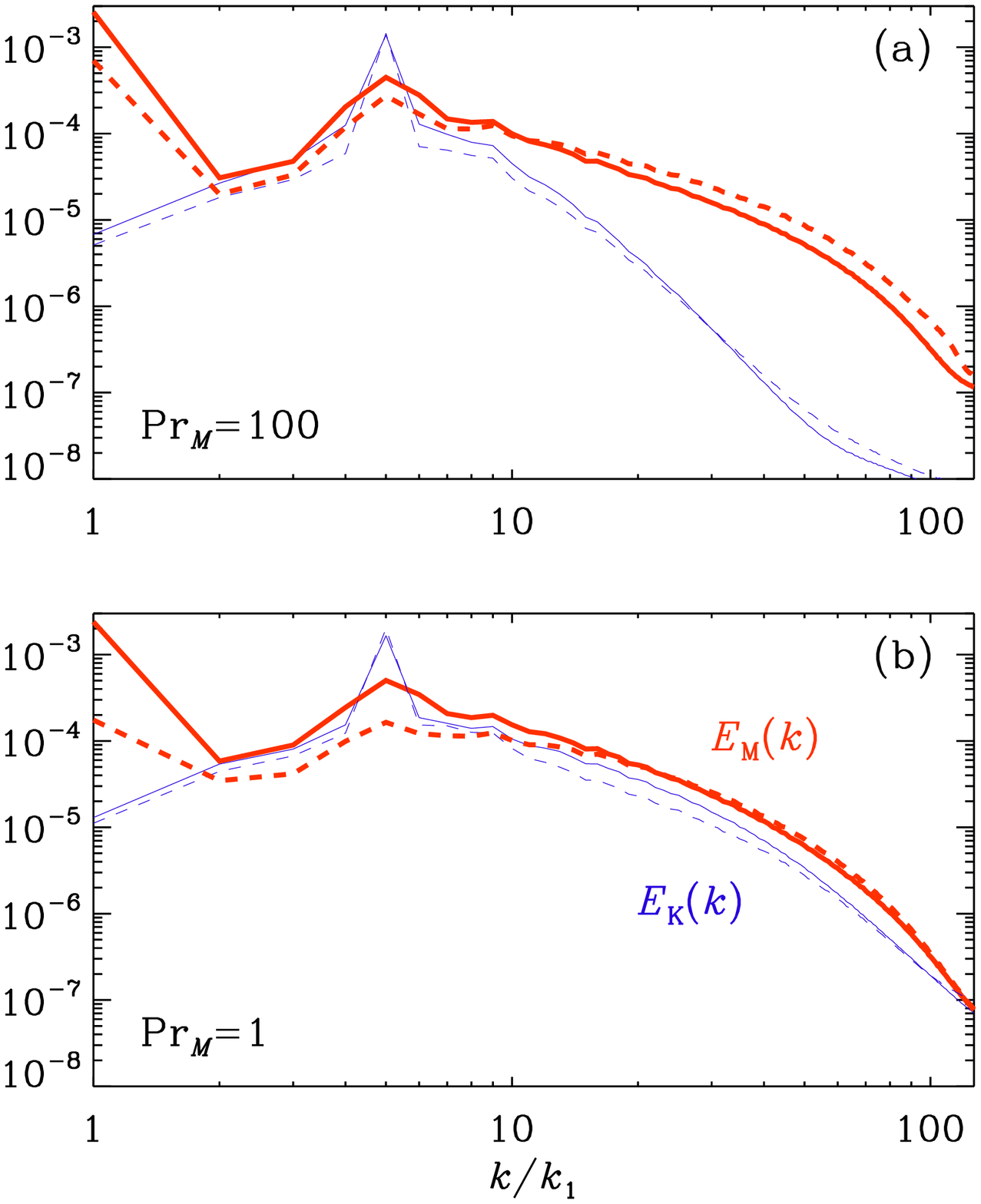}
\end{center}\caption[]{
(Color online)
Comparison of kinetic and magnetic energy spectra for
$\Pm=100$ (upper panel) and $\Pm=1$ (lower panel) for
$\sigma=0.2$ (solid lines) and $0.12$ (dashed lines).
Magnetic energy spectra are shown as thick red lines
while kinetic energy spectra are shown as thin blue lines.
}\label{fig: pspec_comp}
\end{figure}

In \Fig{fig: Bsat_Ca_Pm100} we show results both for $\Pm=100$ and 1.
We discuss first runs for $\Pm=100$
at different values of $\epsf$ and $\Rm$ being either 80, 200, or 600.
Most importantly, it turns out that the critical value for LSD onset
is not much changed.
An extrapolation suggests now $\Cacrit\approx0.9$ instead of 1.
Furthermore, the dependence of $\bra{\meanBB^2}/\Beq^2$ on $\Ca$ is the
same for all three values of $\Rm$, and so $\Cacrit$ is independent of $\Rm$.
However, the values of $\bra{\meanBB^2}/\Beq^2$ are now systematically
above the theoretically expected values.
This discrepancy with the theory can be easily explained by arguing that
the relevant value of $\Beq$ has been underestimated in the large $\Pm$ cases.
Looking at the power spectrum of the high $\Pm$ simulations in
\Fig{fig: pspec_comp}(a), we see that the kinetic energy is indeed
subdominant and does not provide a good estimate of the magnetic energy
of the small-scale field $\bra{\bb^2}/2\mu_0$.
By contrast, for $\Pm=1$, the magnetic and kinetic energy spectra are
similar at all scales except near $k=k_1$; see \Fig{fig: pspec_comp}(b).
The slight super-equipartition for $k>\kf$ is also typical of
a SSD \cite{Haugen_etal2004PhRvE}.

\begin{figure}
\begin{center}
\includegraphics[width=\columnwidth]{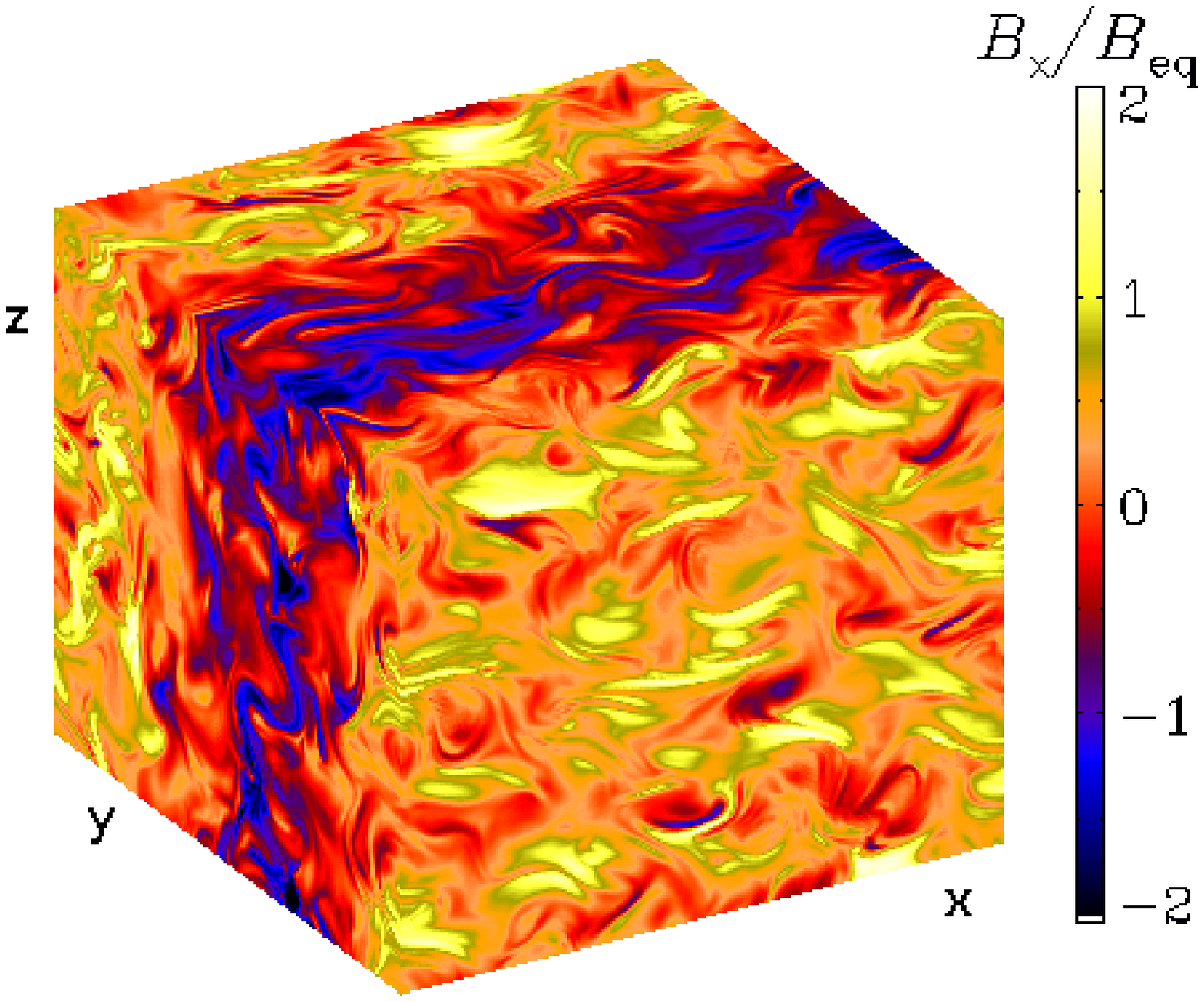}
\end{center}\caption[]{
(Color online)
Visualization of $B_x$ on the periphery of the domain for $\Pm=100$ after
resistive saturation.
}\label{fig: 256sig02_Pm100a}
\end{figure}

A visualization of the magnetic field for $\Pm=100$ is given in
\Fig{fig: 256sig02_Pm100a}, where we show $B_x$ on the periphery
of the computational domain.
The magnetic field has now clearly strong gradients locally, while
still being otherwise dominated by a large-scale component at $k=k_1$.
In this case the large-scale field shows variations only in the $y$ direction
and is of the form
\EQ
\meanBB=(\sin k_1y, 0, \cos k_1y)\,B_{\rm sat}.
\EN
This field has negative magnetic helicity, so
$\meanJJ\cdot\meanBB=-k_1\meanBB^2$, as expected
for a forcing function with negative helicity.

We have argued that the reason for the larger values in the graph of
$\bra{\meanBB^2}$ versus $\Ca$ is related to $\Beq$ being underestimated
for large values of $\Pm$.
To confirm this, we now consider calculations with $\Pm=1$,
different values of $\epsf$ and $\Rm$ (from 168 to 745),
and fixed scale separation ratio $\kf/k_1=5$.
We see in \Fig{fig: Bsat_Ca_Pm100} that the values are now indeed smaller.
An extrapolation would suggest that $\Cacrit$ is now above 1,
but this may not be significant given the uncertainties associated with being
so close to the critical value of $\epsf$.

\begin{figure}[t!]\begin{center}
\includegraphics[width=\columnwidth]{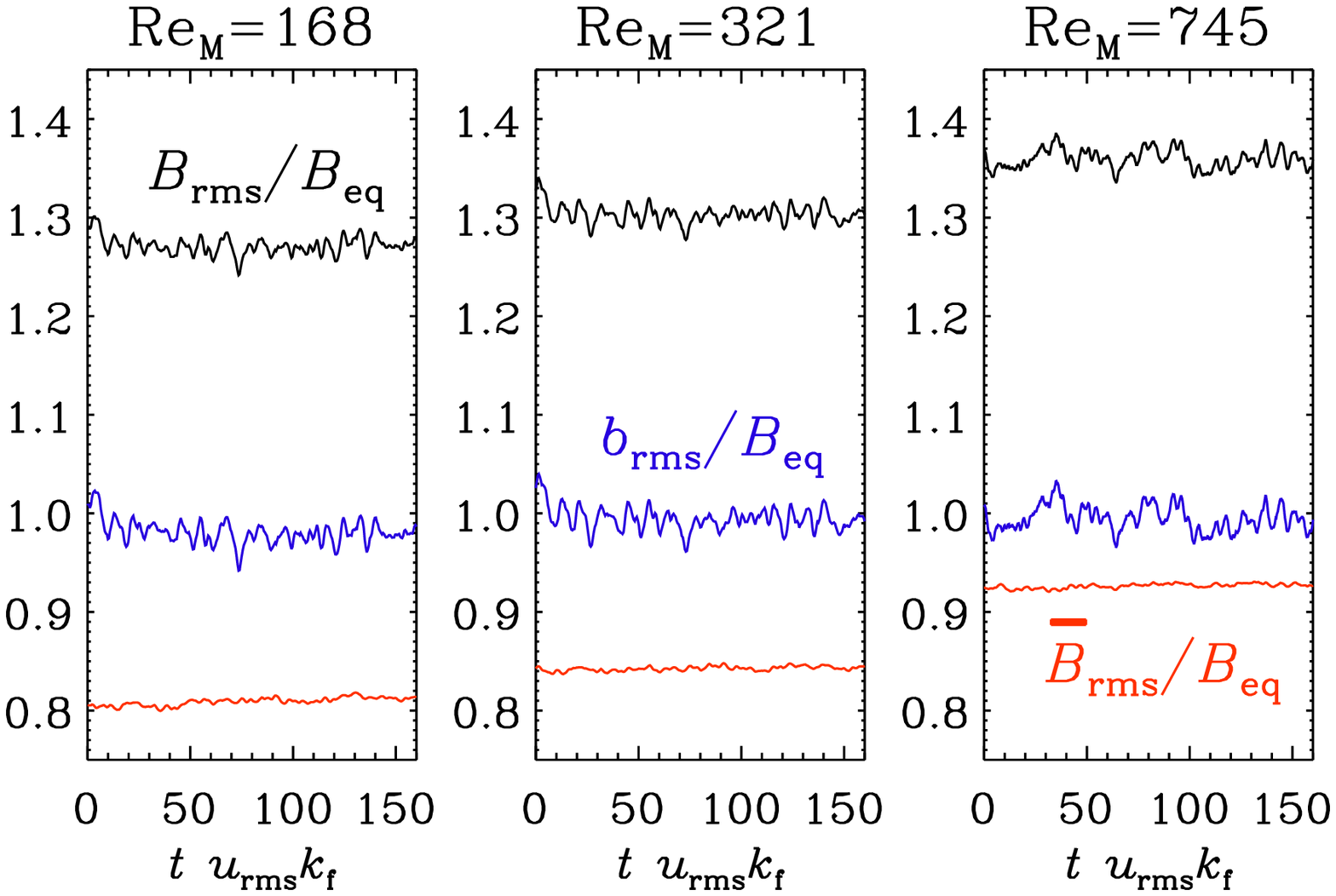}
\end{center}\caption[]{
(Color online)
Evolution of total magnetic field ($B_{\rm rms}$, upper black line),
small-scale magnetic field ($b_{\rm rms}$, blue in the middle),
and large-scale magnetic field ($\meanB_{\rm rms}$, lower red line)
for three values of $\Rm$ over a time stretch of 160 turnover times.
}\label{pncomp_flucts}\end{figure}

LSDs of the type of an $\alpha^2$ dynamo only become apparent in the
late saturation of the dynamo \cite{BrandInverseCascade2001}.
This is especially true in the case of large values of $\Rm$ when the
mean field develops its full strength while the rms value of the
small-scale field remains approximately unchanged as $\Rm$ increases;
see \Fig{pncomp_flucts}.
Note also that the level of fluctuations of both small-scale and
large-scale magnetic fields remains approximately similar for different
values of $\Rm$.
This also shows that the emergence of SSD action does not have any
noticeable effect on the LSD.

\begin{figure}[t!]\begin{center}
\includegraphics[width=\columnwidth]{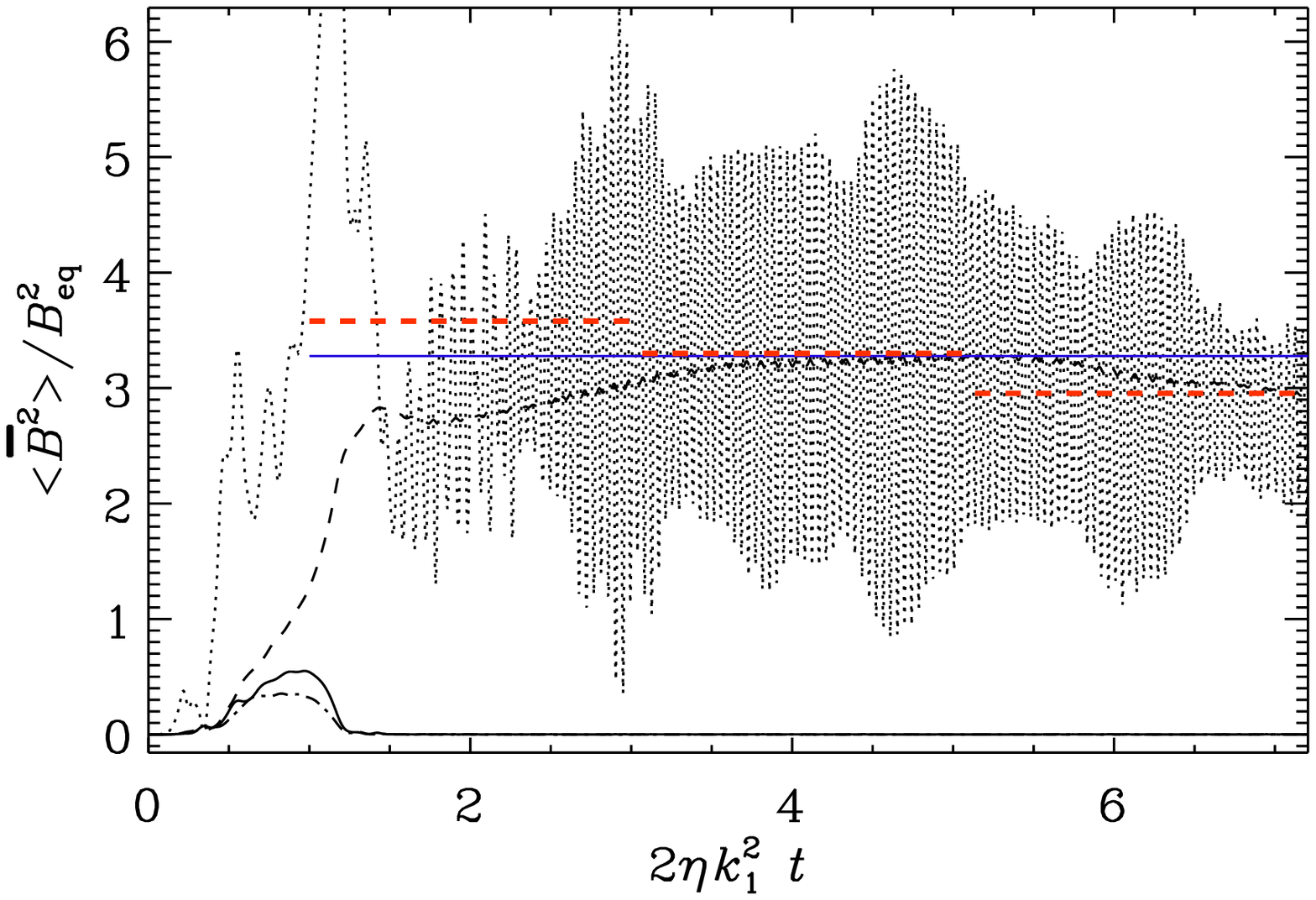}
\end{center}\caption[]{
(Color online)
Similar to \Fig{psat}, but for time-dependent ABC-flow driving.
As in \Fig{psat}, we have here $\kf/k_1=15$ and $\Rm\approx6$.
}\label{psat_t10dep128eta5em3k15a}\end{figure}

\subsection{ABC-flow forcing}
\label{Sec: ABC}

In this paper we have used the fact that the saturation field strength
is described by \Eq{eq: Bsat MF}.
While this is indeed well obeyed for our randomly driven flows, this does
not seem to be the case for turbulence driven by ABC-flow forcing.
We demonstrate this by considering a case that is similar to that shown
in \Fig{psat}, where $\Rm\approx6$ in the saturated state.
We thus use \Eq{ABC} with $\sigma=\theta_0=1$ and $\kf/k_1=15$.
The {\em kinematic} flow velocity reaches an equilibrium rms velocity of
$U_0=f_0/(\nu\kf^2)$.
The magnetic Reynolds number based on this velocity is $U_0/(\eta\kf)$,
which is chosen to be 13, so that during {\em saturation} the resulting value
of $\Rm$ is about 6, just as in \Fig{psat}.
For the $x$, $y$, and $z$ components
we take different forcing frequencies such that $\omega_i/(k_1U_0)$
is 10, 11, and 9 for $i=1$, 2, and 3, respectively.
These values correspond approximately to the inverse correlation times
used in Ref.~\cite{GrahamBlackmanMinuteHelicity12}.
The result is shown in \Fig{psat_t10dep128eta5em3k15a}.
It turns out that the magnetic field grows initially as expected, based
on \Eq{eq: sat}, but then the final saturation phase is cut short below
$B_{\rm sat}^2/\Beq^2\approx3$ rather than the value 12 found with
random wave forcing.
This is reminiscent of inhomogeneous dynamos in which magnetic helicity
fluxes operate.
In homogeneous systems, however, magnetic helicity flux divergences have
only been seen if there is also shear \cite{HubbardBrandenb2012}.
In any case, the present behavior is unexpected and suggests that the
effective value of $\Ca$ is reduced.
Using the test-field method \cite{Schrinner07,BRS08}, we have confirmed
that the actual value of $\Ca$ is not reduced.
The dynamo is therefore excited, but the value implied for the effective
helicity is reduced.

\begin{figure}[t!]\begin{center}
\includegraphics[width=\columnwidth]{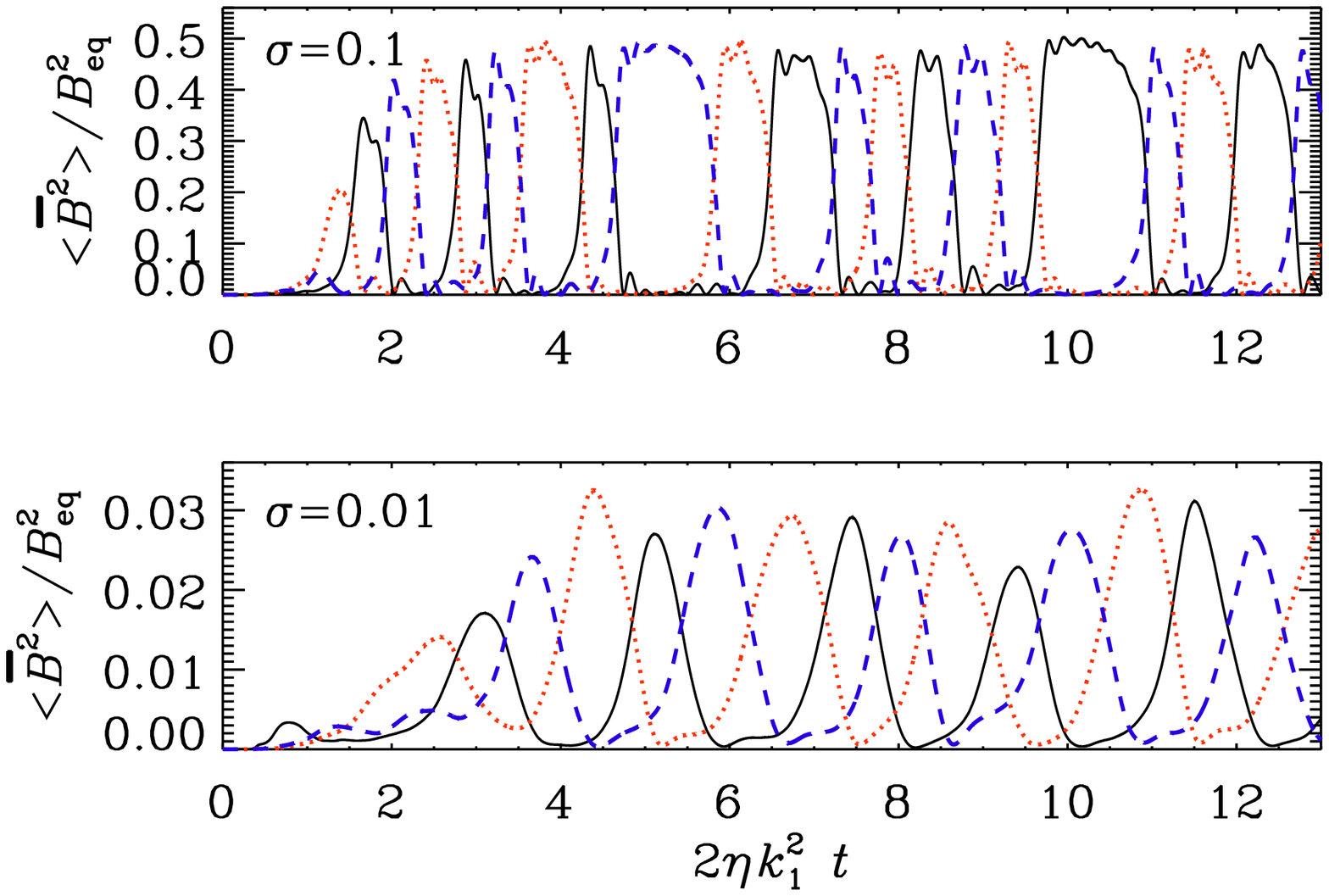}
\end{center}\caption[]{
(Color online)
Dependence of the normalized $\bra{\meanBB^2}$ for different planar
averages: $yz$ (black), $xz$ (red, dotted), and $xy$ (blue, dashed),
for $\sigma=0.1$ (upper panel) and $\sigma=0.01$ (lower panel).
}\label{pcyc_comp}\end{figure}

Another possibility is that, especially for small values of $\sigma$,
the ABC-flow has nongeneric dynamo properties that emulate aspects
of large-scale dynamos.
An example is shown in \Fig{pcyc_comp} where we plot the time evolution
of all three planar averages ($yz$, $xz$, and $xy$).
Even for $\sigma=0.01$, large-scale magnetic fields are still excited,
but the field orientation changes periodically on a timescale of
1--2 diffusion times.
This is obviously a fascinating topic for further research,
but it is unrelated to our main question regarding the minimal helicity
of generic turbulent dynamos.
It might indeed be an example of so-called incoherent $\alpha$ effect
dynamos \cite{VB97} that have recently attracted increased interest
\cite{Pro07,Hei11,MB12}.

The main point of this section is to emphasize the limited usefulness of
ABC-flow dynamos.
Another such example are dynamos driven by the Galloway-Proctor flow,
which also has a number of peculiar features; see Ref.~\cite{RB09}.

\section{Conclusions}

In this paper we have studied the simplest possible LSD and have
investigated the dependence of its saturation amplitude
on the amount of kinetic helicity in the system.
We recall that the case of a periodic domain has already been
investigated in some detail \cite{FieldBlackman2002ApJ,Subramanian2002},
and that theoretical predictions in the case with shear
\cite{BlackmanBrandenb2002ApJ} have been verified numerically
for fractional helicities \cite{KapylaBrandenb2009ApJ}.
Yet the issue has now attracted new interest in view of recent results
suggesting that, in the limit of large scale separation, the amount of
kinetic helicity needed to drive the LSD might actually be much smaller
than what earlier calculations have suggested
\cite{GrahamBlackmanMinuteHelicity12}.
This was surprising given the earlier confirmations of the theory.
As explained above, the reason for the conflicting earlier results
may be the fact that the LSD cannot be safely isolated in the linear
regime, because it will be dominated by the SSD or,
in the case of the ABC-flow dynamo, by some other kind of dynamo
that is not due to the $\alpha$ effect.
Furthermore, as already alluded to in the introduction, there can
be solutions with long-range correlations that could mimic those
that are not due to the $\alpha$ effect.
Within the framework of the Kazantsev model \cite{Kaz68},
the solutions to the resulting Schr\"odinger-type equation can be described
as bound states.
The addition of kinetic helicity leads to new solutions with long-range
correlations as a result of tunneling from the SSD
solutions \citep{S99,BS00,BCR05}.
Indeed, it has been clear for some time that large-scale magnetic fields
of the type of an $\alpha^2$ dynamo become only apparent in the late
saturation of the dynamo \cite{BrandInverseCascade2001}.
This is especially true for the case of large values of $\Rm$ when the
mean field develops its full strength while the rms value of the
\blue{
small-scale field due to SSD action remains approximately unchanged
as $\Rm$ increases; see \Fig{pncomp_flucts}.
}

While there will always remain some uncertainty regarding the application
to the much more extreme astrophysical parameter regime, we can now
rule out the possibility of surprising effects within certain limits
of $\Rm$ and $\Rey$ below 740, and scale separation ratios below 80.
In stars and galaxies, the scale separation ratio is difficult to estimate,
but it is hardly above the largest value considered here.
This ratio is largest in the top layers of the solar convection zone
where the correlation length of the turbulence is short ($1\Mm$) compared
with the spatial extent of the system ($100\Mm$).

Of course, the magnetic Reynolds numbers in the Sun and in galaxies are much
larger than what will ever be possible to simulate.
Nevertheless, the results presented here show very little dependence of the
critical value of $\Ca$ on $\Rm$.
For $\Pm=1$, for example, we find $\Cacrit=1.2$ for $\Rm\approx6$ and
$\Cacrit=1.5$ for $\Rm\approx600$.
On the other hand, for larger values of $\Pm$, the value of $\Cacrit$
can drop below unity ($\Cacrit=0.9$ for $\Pm=100$).
While these changes of $\Cacrit$ are theoretically not well understood,
it seems clear that they are small and do not provide support
for an entirely different scaling law, as anticipated in recent work
\cite{GrahamBlackmanMinuteHelicity12}.

\acknowledgments
The difference between ABC-flows and random plane wave-driven flows
was noted during the Nordita Code Comparison Workshop in August 2012,
organized by Chi-kwan Chan, whom we thank for his efforts.
The authors thank Eric Blackman and Pablo Mininni for useful discussions,
and two anonymous referees for their comments and suggestions.
This work was supported in part by the European Research Council
under the AstroDyn Research Project 227952
and the Swedish Research Council Grant No.\ 621-2007-4064.
Computing resources have been provided by the
Swedish National Allocations Committee at the Center for
Parallel Computers at the Royal Institute of Technology in Stockholm
and Iceland, as well as by the Carnegie Mellon University Supercomputer Center.

\bibliography{references}

\begin{thebibliography}{10}

\bibitem{Parker1979book}
E.~N. {Parker}.
\newblock {\em {Cosmical magnetic fields: Their origin and their activity}}.
\newblock Oxford, Clarendon Press; New York, Oxford University Press, p.\ 858,
  1979.

\bibitem{MoffattBook1978}
H.~K. Moffatt.
\newblock {\em {Magnetic field generation in electrically conducting fluids}}.
\newblock Camb. Univ. Press, 1978.

\bibitem{Radler1980book}
F.~{Krause} and {K.-H.} {R\"adler}.
\newblock {\em {Mean-field magnetohydrodynamics and dynamo theory}}.
\newblock Oxford, Pergamon Press, Ltd., p.~271, 1980.

\bibitem{Seehafer1996}
N.~{Seehafer}.
\newblock {\em \pre}, 53:1283--1286, 1996.

\bibitem{Ji1999}
H.~{Ji}.
\newblock {\em Phys. Rev. Lett.}, 83:3198--3201, 1999.

\bibitem{Frisch-Pouquet-Leorat-1975-JFluidMech}
U.~Frisch, A.~Pouquet, J.~Leorat, and A.~Mazure.
\newblock {\em J. Fluid Mech.}, 68:769--778, 1975.

\bibitem{BlackmanField2000a}
E.~G. {Blackman} and G.~B. {Field}.
\newblock {\em \apj}, 534:984--988, 2000.

\bibitem{BlackmanField2000b}
E.~G. {Blackman} and G.~B. {Field}.
\newblock {\em \mnras}, 318:724--732, 2000.

\bibitem{Kleeorin_etal2000}
N.~{Kleeorin}, D.~{Moss}, I.~{Rogachevskii}, and D.~{Sokoloff}.
\newblock {\em Astron. Astrophys.}, 361:L5--L8, 2000.

\bibitem{ssHelLoss09}
A.~{Brandenburg}, S.~{Candelaresi}, and P.~{Chatterjee}.
\newblock {\em \mnras}, 398:1414--1422, 2009.

\bibitem{GrahamBlackmanMinuteHelicity12}
J.~{Pietarila Graham}, E.~G. {Blackman}, P.~D. {Mininni}, and A.~{Pouquet}.
\newblock {\em \pre}, 85:066406, 2012.

\bibitem{Brandenburg2009ApJ}
A.~{Brandenburg}.
\newblock {\em \apj}, 697:1206--1213, 2009.

\bibitem{Schekochihin_etal2004ApJ}
A.~A. {Schekochihin}, S.~C. {Cowley}, S.~F. {Taylor}, J.~L. {Maron}, and J.~C.
  {McWilliams}.
\newblock {\em \apj}, 612:276--307, 2004.

\bibitem{Haugen_etal2004PhRvE}
N.~E.~L. {Haugen}, A.~{Brandenburg}, and W.~{Dobler}.
\newblock {\em \pre}, 70:016308, 2004.

\bibitem{BrandInverseCascade2001}
A.~{Brandenburg}.
\newblock {\em \apj}, 550:824--840, 2001.

\bibitem{BlackmanBrandenb2002ApJ}
E.~G. {Blackman} and A.~{Brandenburg}.
\newblock {\em \apj}, 579:359--373, 2002.

\bibitem{KapylaBrandenb2009ApJ}
P.~J. {K{\"a}pyl{\"a}} and A.~{Brandenburg}.
\newblock {\em \apj}, 699:1059--1066, 2009.

\bibitem{CH09}
F.~{Cattaneo} and D.~W. {Hughes}.
\newblock {\em \mnras}, 395:L48--L51, 2009.

\bibitem{BRRS08}
A.~{Brandenburg}, K.-H. {R{\"a}dler}, M.~{Rheinhardt}, and K.~{Subramanian}.
\newblock {\em \apjl}, 687:L49--L52, 2008.

\bibitem{BCR05}
S.~{Boldyrev}, F.~{Cattaneo}, and R.~{Rosner}.
\newblock {\em \prl}, 95:255001, 2005.

\bibitem{Kaz68}
A.~P. {Kazantsev}.
\newblock {\em Sov.\ J.\ Exp.\ Theor.\ Phys.}, 26:1031, 1968.

\bibitem{S99}
K.~{Subramanian}.
\newblock {\em \prl}, 83:2957--2960, 1999.

\bibitem{BS00}
A.~{Brandenburg} and K.~{Subramanian}.
\newblock {\em \aap}, 361:L33--L36, 2000.

\bibitem{Gal12}
D.~{Galloway}.
\newblock {\em Geophys.\ Astrophys.\ Fluid Dynam.}, 106:450--467, 2012.

\bibitem{Galanti_etal1992GApFD}
B.~{Galanti}, P.~L. {Sulem}, and A.~{Pouquet}.
\newblock {\em Geophys.\ Astrophys.\ Fluid Dynam.}, 66:183--208, 1992.

\bibitem{Kleeorin_etal2009PhRvE}
N.~{Kleeorin}, I.~{Rogachevskii}, D.~{Sokoloff}, and D.~{Tomin}.
\newblock {\em \pre}, 79:046302, 2009.

\bibitem{PouquetFrischLeorat1976JFM}
A.~{Pouquet}, U.~{Frisch}, and J.~{L\'eorat}.
\newblock {\em J. Fluid Mech.}, 77:321--354, 1976.

\bibitem{RR07}
K.-H. {R{\"a}dler} and M.~{Rheinhardt}.
\newblock {\em Geophys.\ Astrophys.\ Fluid Dynam.}, 101:117--154, 2007.

\bibitem{FieldBlackman2002ApJ}
G.~B. {Field} and E.~G. {Blackman}.
\newblock {\em \apj}, 572:685--692, 2002.

\bibitem{BS07}
A.~{Brandenburg} and K.~{Subramanian}.
\newblock {\em Astron.\ Nachr.}, 328:507, 2007.

\bibitem{SSB07}
S.~{Sur}, K.~{Subramanian}, and A.~{Brandenburg}.
\newblock {\em \mnras}, 376:1238--1250, 2007.

\bibitem{BKM04}
A.~{Brandenburg}, P.~J. {K{\"a}pyl{\"a}}, and A.~{Mohammed}.
\newblock {\em Phys.\ Fluids}, 16:1020--1027, 2004.

\bibitem{Sur_etal2008}
S.~{Sur}, A.~{Brandenburg}, and K.~{Subramanian}.
\newblock {\em \mnras}, 385:L15--L19, 2008.

\bibitem{KRP94}
L.~L. {Kitchatinov}, V.~V. {Pipin}, and G.~{Ruediger}.
\newblock {\em Astron.\ Nachr.}, 315:157--170, 1994.

\bibitem{RK01}
I.~{Rogachevskii} and N.~{Kleeorin}.
\newblock {\em \pre}, 64:056307, 2001.

\bibitem{BP12}
A.~{Brandenburg} and A.~{Petrosyan}.
\newblock {\em Astron. Nachr.}, 333:195, 2012.

\bibitem{Batchelor1953}
G.~K. {Batchelor}.
\newblock {\em {The Theory of Homogeneous Turbulence}}.
\newblock Cambridge: Cambridge University Press, 1953.

\bibitem{Brandenburg2011AN}
A.~{Brandenburg}.
\newblock {\em Astron. Nachr.}, 332:51, 2011.

\bibitem{HubbardBrandenb2012}
A.~{Hubbard} and A.~{Brandenburg}.
\newblock {\em \apj}, 748:51, 2012.

\bibitem{Schrinner07}
M.~{Schrinner}, K.-H. {R{\"a}dler}, D.~{Schmitt}, M.~{Rheinhardt}, and U.~R.
  {Christensen}.
\newblock {\em Geophys.\ Astrophys.\ Fluid Dynam.}, 101:81--116, 2007.

\bibitem{BRS08}
A.~{Brandenburg}, K.-H. {R{\"a}dler}, and M.~{Schrinner}.
\newblock {\em \aap}, 482:739--746, 2008.

\bibitem{VB97}
E.~T. {Vishniac} and A.~{Brandenburg}.
\newblock {\em \apj}, 475:263, 1997.

\bibitem{Pro07}
M.~R.~E. {Proctor}.
\newblock {\em \mnras}, 382:L39--L42, 2007.

\bibitem{Hei11}
T.~{Heinemann}, J.~C. {McWilliams}, and A.~A. {Schekochihin}.
\newblock {\em \prl}, 107:255004, 2011.

\bibitem{MB12}
D.~{Mitra} and A.~{Brandenburg}.
\newblock {\em \mnras}, 420:2170--2177, 2012.

\bibitem{RB09}
K.-H. {R{\"a}dler} and A.~{Brandenburg}.
\newblock {\em \mnras}, 393:113--125, 2009.

\bibitem{Subramanian2002}
K.~{Subramanian}.
\newblock {\em Bull. Astron. Soc. India}, 30:715--721, 2002.

\end{thebibliography}

\end{document}